%% file: sample-manuscript.tex
\newcommand {\ed}[1]{{\color{black}{#1}}}
\newcommand {\edit}[1]{{\color{black}{#1}}}
\documentclass[manuscript,screen]{acmart}
\usepackage{csquotes}
\usepackage{enumitem}
\usepackage[normalem]{ulem}
\useunder{\uline}{\ul}{}
\usepackage{colortbl}
\usepackage{listings}
\usepackage{xcolor}

\definecolor{darkgreen}{rgb}{0.0, 0.5, 0.0}
\definecolor{darkred}{rgb}{0.8, 0.0, 0.1}

\newcommand{\inlinecode}[2]{\lstinline[language=#1]{#2}}

\lstdefinelanguage{text}{
  basicstyle=\ttfamily,
  identifierstyle=\color{black},
  keywordstyle=\color{blue},
  stringstyle=\color{red},
  commentstyle=\color{green},
}

\lstdefinelanguage{python}{
  keywords={pandas, pd, numba, scipy, ultralytics, cv2, matplotlib},
  keywordstyle=\color{blue},
  ndkeywords={from, import, as},
  ndkeywordstyle=\color{darkgreen},
  identifierstyle=\color{black},
  sensitive=false,
  comment=[l]{//},
  morecomment=[s]{/*}{*/},
  commentstyle=\color{green}\ttfamily,
  stringstyle=\color{darkred}\ttfamily,
  morestring=[b]',
  morestring=[b]"
}

\lstdefinelanguage{javascript}{
  keywords={typeof, new, true, false, catch, function, return, null, catch, switch, var, if, in, while, do, else, case, break},
  keywordstyle=\color{blue},
  ndkeywords={from, import, as},
  ndkeywordstyle=\color{darkgreen},
  identifierstyle=\color{black},
  sensitive=false,
  comment=[l]{//},
  morecomment=[s]{/*}{*/},
  commentstyle=\color{green}\ttfamily,
  stringstyle=\color{darkred}\ttfamily,
  morestring=[b]',
  morestring=[b]"
}

\lstdefinelanguage{jsx}[]{javascript}{
  morekeywords={typeof, new, true, false, catch, function, return, null, catch, switch, var, if, in, while, do, else, case, break, export, boolean, throw, implements, import, this, from},
  otherkeywords={< >}
}

\lstdefinelanguage{typescript}[]{javascript}{
  morekeywords={interface, implements, readonly, any, void, public, private, protected},
}

\lstdefinelanguage{tsx}[]{typescript}{
  morekeywords={< >}
}

\AtBeginDocument{%
  \providecommand\BibTeX{{%
    \normalfont B\kern-0.5em{\scshape i\kern-0.25em b}\kern-0.8em\TeX}}}

\setcopyright{acmlicensed}
\copyrightyear{2024}
\acmYear{2024}
\acmDOI{XXXXXXX.XXXXXXX}

\acmConference[Conference acronym 'XX]{Make sure to enter the correct
  conference title from your rights confirmation emai}{June 03--05,
  2018}{Woodstock, NY}

\begin{document}

\title[\textit{Hug Reports}: Supporting Expression of Appreciation \\ between Users and Contributors of Open Source Software Packages]{\textit{Hug Reports}: Supporting Expression of Appreciation between Users and Contributors of Open Source Software Packages}

\author{Pranav Khadpe}
\authornotemark[1]
\email{pkhadpe@cs.cmu.edu}
\affiliation{%
  \institution{Carnegie Mellon University}
  \city{Pittsburgh}
  \state{Pennsylvania}
  \country{USA}
}

\author{Olivia Xu}
\authornote{Both authors contributed equally to this research.}
\email{okx@andrew.cmu.edu}
\affiliation{%
  \institution{Carnegie Mellon University}
  \city{Pittsburgh}
  \state{Pennsylvania}
  \country{USA}
}

\author{Geoff Kaufman}
\email{gfk@cs.cmu.edu}
\affiliation{%
  \institution{Carnegie Mellon University}
  \city{Pittsburgh}
  \state{Pennsylvania}
  \country{USA}
  }

\author{Chinmay Kulkarni}
\email{chinmay.kulkarni@emory.edu}
\affiliation{%
  \institution{Emory University}
  \city{Atlanta}
  \state{Georgia}
  \country{USA}
}

\renewcommand{\shortauthors}{Khadpe and Xu, et al.}

\begin{abstract}
\input{sections/abstract}
\end{abstract}



\begin{teaserfigure}
    \centering
    \includegraphics[width=\textwidth]{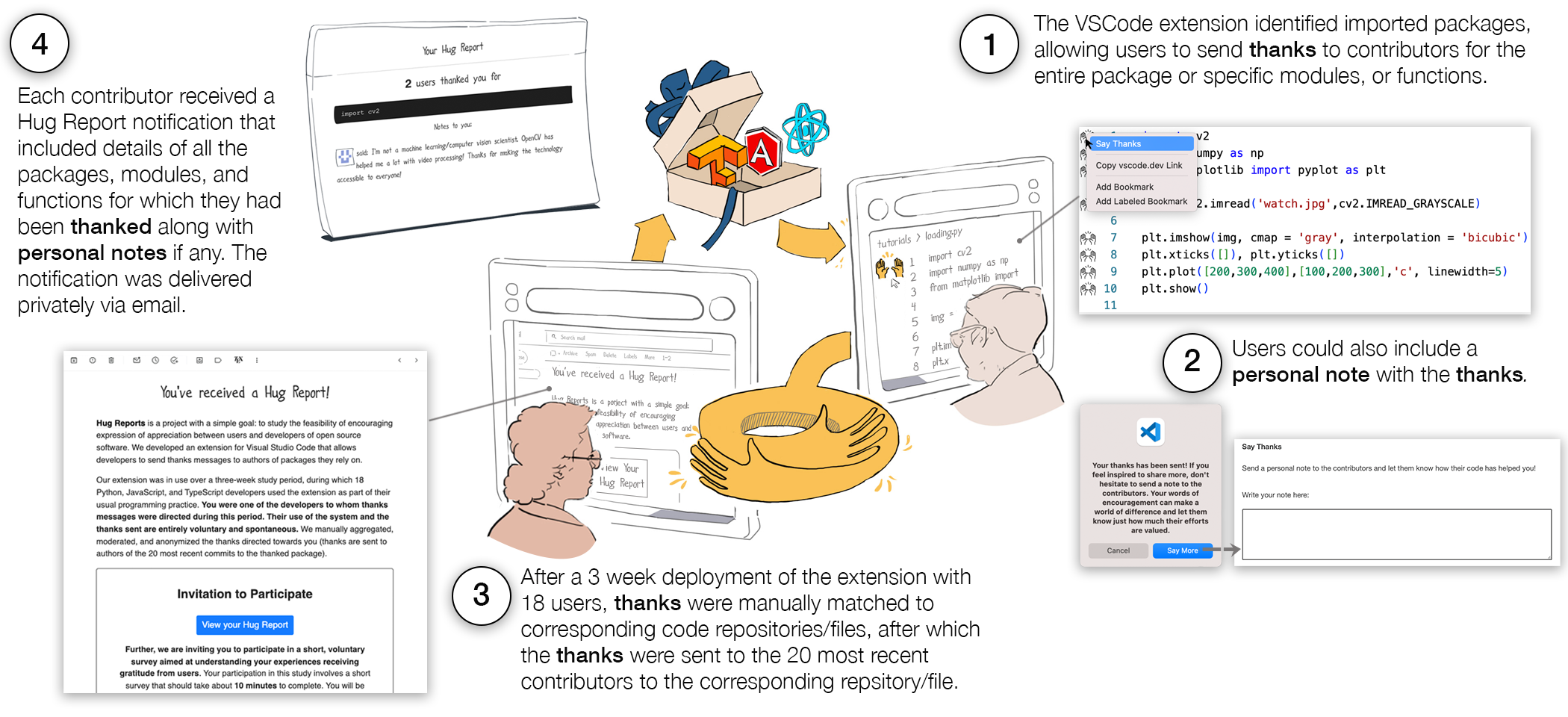}
    \caption{In this paper, we describe a field trial of the Hug Reports \ed{technology probe}. The probe consists of two main components: an extension for the Visual Studio Code editor, and a notification delivered via email. (1) The extension detects imported Python and JavaScript packages in the current file and renders a button on each line that interfaces with an imported package. Clicking the button on any line, logs a \textit{thanks}, and the code on the line specifies the object of the user's appreciation: they can express appreciation for the entire package or specific modules, or functions. (2) With each \textit{thanks}, users can also include a \textit{personal note}. (3) After a 3 week deployment of the extension with 18 users, \textit{thanks} were manually matched to the corresponding source code repository/file, from where we identified the 20 most recent contributors of the thanked work. (4) These contributors were contacted via an email containing the Hug Report notification which displayed the work for which they had been \textit{thanked}. The email also contained a description of our project and an invitation to participate in the study.}
    \label{fig:teaserfig}
\end{teaserfigure}

\maketitle
\section{Introduction}
\input{sections/introduction}
\section{Background and Related Work}
\input{sections/related}
\section{Hug Reports Concept Proposal and Study Overview}
\input{sections/concept-and-overview}

\section{Study}
\input{sections/study}
\section{Findings}
\input{sections/findings}
\section{Discussion}
\input{sections/discussion}
\section{Conclusion}
\input{sections/conclusions}

\bibliographystyle{ACM-Reference-Format}
\bibliography{bibliography}

\appendix

\input{sections/appendix}

\end{document}

%% file: sections/abstract.tex
Contributors to open source software packages often describe feeling discouraged by the lack of positive feedback from users. 
This paper describes a \ed{technology probe}, Hug Reports, that provides users a communication affordance within their code editors, through which users can convey appreciation to contributors of packages they use. In our field study, 18 users interacted with the \ed{probe} for 3 weeks, resulting in messages of appreciation to 550 contributors, 26 of whom participated in subsequent research. \ed{Our findings show how locating a communication affordance within the code editor, and allowing users to express appreciation in terms of the abstractions they are exposed to (packages, modules, functions), can support exchanges of appreciation that are meaningful to users and contributors. Findings also revealed the moments in which users expressed appreciation, the two meanings that appreciation took on---as a measure of utility and as an act of expressive communication---and how contributors' reactions to appreciation were influenced by their perceived level of contribution. Based on these findings,  we discuss opportunities and challenges for designing appreciation systems for open source in particular, and peer production communities more generally.}

%% file: sections/introduction.tex
\begin{displayquote}
{\textit{Users are far more likely to reach out when they have a complaint. If everything works great, they tend to stay silent. It can be discouraging to see a growing list of issues without the positive feedback showing how your contributions are making a difference.}}\par
\raggedleft
—Abby Cabunoc Mayes, Maintaining Balance for Open Source Maintainers~\cite{Mayes}
\end{displayquote}

\begin{displayquote}
{\textit{In a sense, these GitHub notifications are a constant stream of negativity about your projects. Nobody opens an issue or a pull request when they’re satisfied with your work. They only do so when they’ve found something lacking. Even if you only spend a little bit of time reading through these notifications, it can be mentally and emotionally exhausting.}}\par
\raggedleft
—Nolan Lawson, What it Feels Like to be an Open-Source Maintainer~\cite{Lawson_2017}
\end{displayquote}


Contributors to open source software packages rarely receive positive feedback from users. Today, many open source packages have become vital digital infrastructure that governments, private companies, and individual developers rely on~\cite{eghbal2016roads}. However, the contributors who develop and maintain these packages---who are often volunteers~\cite{zlotnick_2017_806811, eghbal2020working, champion2021underproduction}---describe how their interactions with those who benefit from their labor tend to be overwhelmingly critical and negative~\cite{miller2022did}. Like the opening quotes, many contributors have talked about the rarity of positive feedback in blog posts, talks, and social media posts, describing how it can feel exhausting and demotivating~\cite{Mayes, Hammer_2021, Lawson_2017, miller2022did}. Lack of positive feedback is often discussed as one of the factors leading to contributor burnout and disengagement~\cite{Mayes}, which have become growing concerns within communities of contributors~\cite{Mayes} and have also been the focus of recent CSCW research~\cite{hsieh2023nip, miller2019people}. The concerns are amplified by the fact that burnout and disengagement increase the risk of critical projects slowing down or even being abandoned~\cite{coelho2017modern}.   

Why is appreciation rare? \ed{We suggest that many barriers to expressing appreciation stem from the fact that \textit{where} users might feel appreciation (in their development environment) and \textit{what} they might feel appreciation towards (a package, its modules, or its functions) is detached from \textit{where} contribution activities occur (social coding platforms like GitHub) and \textit{what} its units are (individual commits or pull requests). This impedes appreciation in several ways. To start with, (1) users must incur effort to navigate to existing communication channels and to locate whom to thank. Channels for communicating with contributors, such as affordances available through social coding platforms, are detached from users’ development environments. So, establishing contact with contributors requires users to leave their development environment, and navigate to communication channels elsewhere. Further, affordances of social coding platforms that allow users to direct interactions towards authors of individual commits or pull requests quickly become ill-suited when a user wants to thank developers of the entire package or its modules. This is because a package or module may combine several commits and pull requests, and contacting authors of each atomic contribution can be prohibitively effortful. Even though social coding platforms provide other channels through which users can notify projects as a whole, (2) these channels de-emphasize appreciation because they were primarily designed to coordinate contribution activities. For instance, the primary channel through which users interact with project contributors on GitHub is by creating ``Issues'', which are intended to ``track ideas, feedback, tasks, or bugs for work on GitHub''. The feature's advertised use (and name) normatively encourages communication around areas of improvement rather than appreciation. Finally, (3) when users work with a package in the development environment, the contributors and their labor are left out of focus. Today, it is possible to discover, download, and use a software package without ever learning the contributors’ identity, much less interacting with them. The impersonal way in which packages are consumed foregrounds the technical capabilities of the software while obscuring contributors' labor, promoting an ignorant or asocial orientation towards contributors~\cite{widder2023dislocated}. Because the labor of contributors is almost forgotten from the development environment, users may fail to feel a sense of personal obligation or appreciation towards contributors.

\subsection{Approach}
As one solution to encourage appreciation, we arrived at a conceptual design proposal for the \textit{Hug Reports} system, a unidirectional communication system that would afford users the ability to select a package currently in use and direct a thanks message towards contributors, from within their code editors. The affordance would allow users to direct thanks towards entire packages (e.g. \inlinecode{python}{matplotlib}), their modules (e.g. \inlinecode{python}{pyplot}), or specific functions within those modules (e.g. \inlinecode{python}{pyplot.scatter}).  Our hope was that locating a communication affordance within the code editor and allowing users to express appreciation in terms of the abstractions they were exposed to (packages, modules, or functions), would lower the effort in communicating appreciation and that the affordance's material presence within the editor would remind users of the contributors behind the packages they use. By explicitly accounting for the manner in which software is ultimately used in the development environment, this approach departs from prior attempts to support appreciation in open source~\cite{overney2020not, zhang2022and}. By capturing appreciation in terms of abstractions of use rather than low-level units of work, this approach is also distinct from other appreciation systems in peer production that capture appreciation towards individual units of contribution~\cite{matiasdiffusion}. Given these differences from pior work, we wanted to investigate the merits of the approach. At the same time, due to our departure from previous approaches, we anticipated that unexpected
cross-cutting constraints and requirements would reveal themselves when an end-to-end system is deployed.

Therefore, in this paper, we describe a field study of the Hug Reports technology probe (Figure \ref{fig:teaserfig}). We developed a barebones version of Hug Reports as a technology probe~\cite{hutchinson2003technology} to investigate: (1) how key decisions of Hug Reports affected the expression of appreciation; (2) how appreciation was felt, expressed, received, and interpreted; and (3) the opportunities that users and contributors identified for further supporting exchanges of appreciation. The probe consists of two main components: an extension for the Visual Studio Code editor through which users can express their appreciation and a notification delivered to contributors via email. The extension renders a button on every line of code that interfaces with an imported Python or JavaScript package, through which users could log a one-bit \textit{thanks} and optionally a \textit{personal note}, corresponding to the package (or its specific modules, classes, or functions) invoked on that line. We deployed the extension for 3 weeks with 18 developers, following which we sent notifications to the 550 contributors whose work was thanked during the deployment. 26 of these contributors participated in our subsequent research activities. 

\subsection{Contributions}
Our study had several key takeaways, including that Hug Reports encouraged appreciation that was meaningful to users and contributors, that appreciation was interpreted both as a measure of utility and as an act of expressive communication, and that contributors' reactions to appreciation were influenced by how much they felt they had contributed to what was thanked. In addition to this, our study revealed patterns in \textit{when} users expressed appreciation. Based on these findings, we discuss the opportunity and limitation of capturing appreciation towards abstractions of use (packages, modules) rather than low-level units of work (individual commits or pull requests), and we discuss opportunities for encouraging appreciation in software development practice.

To summarize, this paper makes the following contributions:
\begin{itemize}
    \item Through the development and deployment of the Hug Reports technology probe, we provide preliminary evidence that locating a communication affordance within the code editor and allowing users to express appreciation in terms of the abstractions they are exposed to, can encourage exchanges of appreciation towards contributors.
    \item We extend prior literature on appreciation in open source by contributing insights into how appreciation is experienced and expressed by users in practice. 
    \item We contribute new design knowledge on designing appreciation systems in peer production by exploring the implications of re-orienting appreciation around abstractions of use.
\end{itemize}

}

%% file: sections/related.tex
In this section, we first describe how open source software packages are developed and distributed, \ed{with a focus on the configuration of interlinking social practices and technical systems involved. Then, by considering the interaction of social and technical factors, we describe the barriers that prevent expression of appreciation. We then discuss how our work departs from prior appreciation systems. Finally, we discuss how the meaningfulness of appreciation can be preserved even as we attempt to lower the effort of expressing appreciation.}
\ed{\subsection{The sociotechnical system within which modern open source packages are produced and used}}
\ed{A central premise of CSCW research is that social practices and technical objects can't be fully understood in isolation; they influence each other~\cite{bowker2014social}. The critical interaction between social and technical factors is captured in the concept of a \textit{sociotechnical system}~\cite{cherns1976principles} where analysis is not at the level of individual technologies but instead considers the broader coherent system of technical objects and human practices. So, to understand barriers to appreciation, we begin by demarcating the sociotechnical system we are designing within. In this work, we focus on open source software projects that are hosted on GitHub\footnote{https://github.com/}, and that are made available to users as packages. Here, we briefly describe how these projects are hosted, developed, and distributed with a focus on both the social practices and the technical objects}.  

Open source software projects produce software that is licensed as ‘open source’, granting future users the rights to use it, study its source code, modify it, and redistribute it, all at no cost~\cite{hippel2003open}. The free distribution and open development practices enable, and often result in, collaborative and social practices of developing and maintaining the software, attracting contributions from distributed developers, many of whom are volunteers~\cite{krishnamurthy2006intrinsic, hippel2003open, benkler2017peer}. For this reason, open source projects have been described as instances of ``peer production''~\cite{benkler2017peer}. Today several open source software projects have become essential digital infrastructure that individual developers as well as commercial firms rely on~\cite{eghbal2016roads, benkler2017peer, geiger2021labor}. Despite growing commercialization and paid development work in open source projects~\cite{germonprez2018eight}, many projects continue to rely heavily on volunteers~\cite{zlotnick_2017_806811, eghbal2020working, champion2021underproduction}.  

With its collaborative and social practices, contemporary open source software development occurs primarily through social coding platforms~\cite{dabbish2012social, dabbish2012leveraging}, most notably GitHub. Social coding platforms provide code hosting capabilities and social features that support collaboration~\cite{dabbish2012social}. This includes features that increase visibility into development activity~\cite{dabbish2012leveraging} as well as channels for communication~\cite{dabbish2012social}. On GitHub, which is the most popular social coding platform~\cite{hsieh2023nip, li2021code}, projects are organized as repositories; a project’s \textbf{repository} contains its source code and hosts the conversations surrounding the project. Permissions to make \textbf{commits} (applying a code patch) to the hosted version of the source code are restricted to those designated as owners or collaborators on the repository. Developers who do not have such permissions, can nonetheless author code patches and work with owners or collaborators who can apply the patches on their behalf. In this work, we use the term `contributor' to mean any developer who has authored a code patch that has eventually been applied to the hosted version, and the term `maintainer' to refer to the subset of contributors who have commit permissions. Any developer interested in keeping track of a project can \textbf{star} the repository. In each repository, developers who are contributors or users can track ideas, feedback, tasks, and bugs by creating \textbf{issues}. When a developer hopes to contribute a code patch to the project, they can open a \textbf{pull request}, a special issue that hosts conversations concerning the contribution. Activities on GitHub leave public traces. This provides visibility into software development activities at the level of actions provided by GitHub, and the underlying Git version control system~\cite{dabbish2012social}.

While development activities take place on GitHub, use often occurs elsewhere. Many projects are part of a larger ``software ecosystem''~\cite{bogart2021and, valiev2018ecosystem}, which Bogart et al. define as ``communities built around
shared programming languages, shared platforms, or shared dependency management tools, allowing developers to create packages that import and build on each others’ functionality''~\cite{bogart2021and}. Within an ecosystem (e.g. npm which is the JavaScript ecosystem or PyPI which is the Python ecosystem), each project is packaged for use, indexed and advertised in a registry, and made available for installation via a package manager~\cite{bogart2021and}, which a user can access through a terminal in their development environment. It is through this distribution channel~\cite{mancinelli2006managing} that many projects are consumed, often from within a development environment.

\subsection{Barriers to expressing appreciation in open source}
Within developer communities~\cite{Mayes, Hammer_2021, Lawson_2017} as well as academic literature~\cite{raman2020stress, miller2022did, hsieh2023nip}, there is growing recognition of how working in open source projects can often feel demotivating and stressful. Open source practitioners have shared experiences~\cite{Hammer_2021, Lawson_2017} about how their interactions with users can feel like a ``constant stream of negativity''~\cite{Lawson_2017}. One reason why interactions can feel overwhelmingly negative is that contributors rarely receive positive feedback and appreciation from the users who benefit from their, often voluntary, labor. Nic Crane, a maintainer of Apache Arrow, suggests: ``We have lots of happy but quiet users''~\cite{Mayes}. With little appreciation, contributors are left facing a stream of user demands and requests, some of which are even aggressive in their tone~\cite{raman2020stress, miller2022did}. In many discussions, the lack of positive feedback and recognition is often discussed as a cause of demotivation and burnout~\cite{Mayes, hsieh2023nip, raman2020stress}.

\ed{We suggest that many barriers to expressing appreciation stem from the fact that \textit{where} users might feel appreciation (in their development environment) and \textit{what} they might feel appreciation towards (a package, its modules, or its functions) is detached from \textit{where} contribution activities occur (GitHub) and \textit{what} its units are (individual commits or pull requests). This impedes appreciation in several ways:

\subsubsection{Users must incur effort to navigate to existing communication channels and to locate whom to thank} Channels for communicating with contributors are available on GitHub, where contribution activities occur. But these are detached from users' development environments, where the packages are used. To contact contributors, users need to leave their development environment, find the right GitHub repository, navigate to GitHub, and then either open an issue or locate the contributor's contact details. Users are more likely to undertake this effort to report issues since improvements can directly benefit them than to express appreciation, from which they might not see any direct benefits. This tendency for negative feedback over positive is also observed in consumer research, showing that customers with bad experiences are more inclined to leave online reviews than those with good experiences~\cite{han2020customer, anderson1998customer}. Further, using GitHub's affordances to discover and contact contributors of individual commits or pull requests is straightforward because contribution activities are organized along those low-level units. But affordances that are oriented towards commits or pull requests quickly become ill-suited when a user wants to thank developers of the entire package or its modules. A module, for instance, may combine several commits and pull requests. So, identifying all relevant contributions, and their authors, can be prohibitively effortful. Because individual commits or pull requests are abstracted away from users when they are working with a package, we suggest it can be beneficial for appreciation systems to be oriented towards the abstractions that users are exposed to---the package or its modules.

\subsubsection{Existing communication channels de-emphasize appreciation because they were primarily designed to coordinate contribution activities} In addition to supporting communication around individual commits and pull requests, GitHub provides some support for communicating with the project as a whole through issues and stars. However, the uptake of these features for communicating appreciation is limited. This is because, in addition to being removed from where users are, these affordances were also designed primarily to coordinate contribution activities, with values of efficiency and productivity in mind.} GitHub ``Issues''\footnote{https://docs.github.com/en/issues/tracking-your-work-with-issues/about-issues}, for instance, are advertised as a way to ``track ideas, feedback, tasks, or bugs for work on GitHub''. To support effective tracking of tasks, projects can further customize the feature to provide users default issue templates for common communication such as bug reporting or feature requests\footnote{https://docs.github.com/en/communities/using-templates-to-encourage-useful-issues-and-pull-requests/configuring-issue-templates-for-your-repository}. Together, the name, advertised purpose, and user defaults normatively discourage appreciation. With hesitation, motivated users still reappropriate GitHub's features to convey appreciation, as visible in issue \#2264 in the react boilerplate project\footnote{https://github.com/react-boilerplate/react-boilerplate/issues/2264}, which is titled: ``I don't know how to thank, and show my appreciation, to the contributors on a good way''. GitHub users have pointed out how the reaction palette available within issues and pull requests lacks emojis to convey the sentiment of ``Thank you''\footnote{https://github.com/orgs/community/discussions/38201}. While stars can be used to convey appreciation, they are also intended to function as a bookmarking tool and a way for the platform to learn user preferences\footnote{https://docs.github.com/en/get-started/exploring-projects-on-github/saving-repositories-with-stars}, which obscures the intention behind `starring'.  
\ed{
\subsubsection{Reduced visibility of contributors' labor in the development environment diminishes appreciation.} In his book ``Exchange and Power in Social Life''~\cite{blau2017exchange}, Blau argues that ``only social exchange tends to engender feelings of personal obligation, gratitude, and trust; purely economic exchange as such does not.'' In the development environment, the contributors and their labor are left out of focus. So, it can be easy to forget that the package relies on the labor of its contributors. The package, then, can seem more like a free economic commodity~\cite{widder2023dislocated} rather than a gift of labor~\cite{Hammer_2021, terranova2000free}. Because the labor of contributors is almost forgotten from the development environment, users may fail to feel a sense of personal obligation or appreciation towards contributors.

To overcome the above barriers to appreciation in open source, our work: (1) attempts to develop a cross-cutting communication channel that brings the ability to contact contributors---for which users presently need to visit and search GitHub---to the development environment, where users work with the packages; (2) attempts to support expression of appreciation in terms of the abstractions users are exposed to---packages, modules, or functions---rather than lower-level units such as individual commits or pull requests; and (3) attempts to provide users with a reminder of contributors' efforts, within the development environment.}
\ed{
\subsection{Appreciation systems in peer production}
\label{prior}
There have been several efforts to develop appreciation systems for open source, as well as other peer production contexts. Here, we identify the ways in which our work departs from these prior efforts. By doing so, we describe how this work extends CSCW literature on designing appreciation systems for the context of open source in particular, and for peer production contexts more generally. Following Spiro et al.~\cite{spiro2016networks}, we use the term ``appreciation systems'' to refer to platforms through which users exchange thanks and praise. Here, we consider ``appreciation'' to include different types of responses to receiving help, where the exchange involves \textit{unspecified} obligations~\cite{blau2017exchange}. This means the person who received help was not required to respond in a specific way beforehand. Examples include appreciation messages, donations, and tips. Although the person may feel an obligation to donate, tip, or say thanks, the exact nature of this obligation is usually not agreed upon in advance~\cite{blau2017exchange, smith2010theory}. Following from this criteria, we do not consider ``bounties'' as appreciation since the reward or compensation is agreed upon in advance.

Our approach departs from prior attempts to support appreciation in open source by explicitly accounting for the manner in which software is ultimately used in the development environment. Prior work has studied several systems through which users of open source software can convey appreciation~\cite{overney2020not, zhang2022and}. This includes donation platforms such as PayPal, Patreon, and OpenCollective which projects may link to from their repositories~\cite{overney2020not}. Similarly, GitHub has the `Sponsors'\cite{shimada2022github} feature through which users of the platform can sponsor individual developers. The `Say Thanks'\footnote{https://github.com/BlitzKraft/saythanks.io} project provides a link that contributors can include to the repository of a project, and that users can visit to send messages of appreciation. However, much like the GitHub features we described in the previous sections, all of these appreciation systems are detached from \textit{where} software use occurs---the development environment. As a result, they present many of the barriers that we described in the previous section. By developing an appreciation system that is fine-tuned to users' development practices, we investigate the potential benefits of such an approach as well as cross-cutting concerns that arise in its implementation. Developing an appreciation system that connects to the users' development environment also alllows us to investigate how appreciation is experienced and expressed by users. This allows us to further extend prior literature because prior studies tend to focus solely on the experiences of the contributors receiving appreciation~\cite{overney2020not, shimada2022github}

Our approach tries to enable users to express appreciation in terms of the abstractions they are exposed to---the package or its modules---rather than lower-level units of contribution such as individual commits or pull requests. This distinguishes it from some existing appreciation systems in peer production that restrict users to expressing appreciation towards individual units of work. Consider Wikipedia's ``Thanks'' system~\cite{matiasdiffusion, goel2019thanks}, through which editors can thank each other. A "thank" link is shown next to each edit in the history view of an article. Clicking the link triggers a notification to the author of the edit. However, having to thank individual edits can be limiting when users want to express appreciation towards a group of edits, a paragraph, or a section. One comment on the talk page for Wikipedia’s “Thanks” system notes: ``Sometimes I'd like to express thanks for a group of edits — for example, when none of them is individually a big deal, but together they're really helpful. Any chance that we could get the chance to issue a single Thanks feature notice for a group of edits?''\footnote{https://en.wikipedia.org/wiki/Help\_talk:Notifications/Thanks/Archive\_2}. Our work adapts \textit{what} users can thank to the form in which they consume the artifact (packages/modules) rather than the unit of production (pull requests or individual commits). We investigate the merits of such adaptation and reveal the constraints and requirements it entails. This allows us to derive implications for designing appreciation systems in other peer production contexts where units of work may be misaligned with \textit{what} users feel appreciation towards. 

\subsection{Lowering effort in expressing appreciation while preserving meaningfulness}
Central to our approach is the idea of lowering the effort for users to express appreciation, by fine-tuning the system to the manner in which software is used in the development environment. Before we can proceed, however, we must address an apparent dilemma: while lowering the effort involved in expressing appreciation may encourage appreciation, it can also undermine its meaningfulness. While low-effort actions can encourage interactions, prior work suggests these interactions can feel limited in the authenticity~\cite{monroy2011computers, zhang2022auggie} and support~\cite{wohn2016affective} they convey. Is lowering effort bound to dilute the meaningfulness of appreciation?  

Prior work offers a resolution to this dilemma by pointing out how some kinds of effort are not considered meaningful. In the context of interpersonal communication, Markopoulos distinguishes between \textit{procedural effort}, which he describes as ``the effort that one needs to expend in order to operate a system''~\cite{markopoulos2009design}  (examples in our context would include logging in, navigating to a repository, finding individual commits) and \textit{personal effort}, which he describes as ``the effort put to attend personally to an individual''~\cite{markopoulos2009design} (which in our context would be composing a thoughtful message). Prior work suggests procedural effort tends not to be valued~\cite{markopoulos2009design, zhang2022auggie} and that it can be minimized to create greater opportunities for personal effort, which is what makes the communication feel special to recipients~\cite{romero2007connecting, markopoulos2009design}. 

In our work, it is specifically the procedural effort that we attempt to lower. Our approach attempts to lower procedural effort by locating a communication affordance within the code editor and by allowing users to express appreciation at the level of packages, modules, or functions rather than individual commits or pull requests. At the same time, we give users control over the amount of personal effort they invest by allowing them to customize their appreciation messages.
}

%% file: sections/concept-and-overview.tex
\ed{\subsection{Concept proposal}}In envisioning solutions that would overcome barriers in expressing appreciation, we arrived at a conceptual design proposal for the \textit{Hug Reports}\footnote{We chose the name to suggest an inversion of the concept of bug reports in software development, which are intended to convey critical feedback.} system, a unidirectional communication system that would afford users the ability to select a package currently in use and direct a ``thanks'' towards contributors from within their code editors. \ed{The affordance would allow users to direct thanks towards entire packages (e.g. \inlinecode{python}{matplotlib}), their modules (e.g. \inlinecode{python}{pyplot}), or specific functions within those modules (e.g. \inlinecode{python}{pyplot.scatter})}. To map the thanked packages to contributors, we planned to use activity traces from GitHub repositories corresponding to the packages to identify the relevant contributors. Finally, we would deliver the ``thanks'' to contributors on behalf of the users. 

\ed{The key decisions in Hug Reports were to: (1) \textbf{lower procedural effort} by locating a communication affordance within the code editor and allowing users to express appreciation at the level of packages, modules, or functions rather than individual commits or pull-requests; and (2) \textbf{provide users a subtle reminder of the contributors} through the material presence of the affordance in the code editor. As we describe in Section \ref{prior}, these decisions differentiate our approach from prior appreciation systems, especially those in open source. So, we wanted to investigate the merits of these key decisions. At the same time, due to our departure from previous approaches, we anticipated that unexpected cross-cutting constraints and requirements would reveal themselves when an end-to-end system is deployed. 

\subsection{Overview of method and research questions}
At this early stage of the design process, then, we required a method to assess the feasibility of the approach and to identify cross-cutting concerns. We wanted to rapidly explore the design space and iterate on our design concept while reducing the risk of developing an end-to-end technical system that users and contributors did not ultimately want So, we chose to use the method of \textit{technology probes}. As per Hutchinson et al. ~\cite{hutchinson2003technology}, technology probes are functioning technological artifacts that balance three goals: \textit{social science}: ``understanding the use and the users''; \textit{engineering}: ``field testing the technology''; and \textit{design}: ``inspiring users to think of new kinds of technology to support their needs''. Technology probes are often used in the design of social applications, to engage participants early in the design process~\cite{jorke2023pearl, sellen2006homenote, leong2023social, khadpe2024discern} and to find out about the ``unknown'' when deployed~\cite{hutchinson2003technology}. So, we chose to develop a barebones version of Hug Reports as a technology probe and deploy it in a field study to address the following research questions:

\edit{
\begin{itemize} [label={}]
\item \textbf{RQ1}---\textit{How do the key decisions of Hug Reports impact expression of appreciation?} To what extent does lowering procedural effort encourage users to express appreciation? To what extent does reminding users of the contributors encourage users to express appreciation? To what extent do contributors find the appreciation meaningful? (This RQ revolves around the probe's goal of ``field testing'' the key decisions.)

\item \textbf{RQ2}---\textit{How is appreciation felt, expressed, received, and interpreted?} How did users express appreciation, in terms of when they felt appreciation, what they felt appreciation towards, and how they articulated it? How were contributors' perceptions of the appreciation affected by what was being appreciated and how appreciation was articulated? (This RQ revolves around the probe's goal of ``understanding the use and users''.)

\item \textbf{RQ3}---\textit{What opportunities do users and contributors envision for supporting appreciation?} (This RQ revolves around the probe's goal of ``inspiring users to think of new kinds of technology to support their needs''.)

\end{itemize}
}
}
In our research activities, we planned to involve two populations of participants: users of open source software packages and contributors to open source software packages. \textbf{Hereafter, we slightly overload terms and refer to participants whom we planned to involve/were involved in their capacity as users, simply as ``users''. Similarly, we refer to participants whom we planned to involve/were involved in their capacity as contributors to packages, simply as ``contributors''.} For the study, we decided to develop realistic versions of the two participant-facing components of the concept: an extension for code editors through which users would express appreciation and the notification system that would convey the appreciation to contributors. 

\ed{\subsection{Considerations in implementing the probe and designing the study} 
}

Implementation of our technology probe and the design of the study were guided by the following factors: 

\textbf{(1) Naturalistic interactions.} Our study prioritized external validity by allowing naturalistic interactions: contributors would only see ``thanks'' sent by actual users and users' ``thanks'' would actually be sent to contributors. Users in the study were informed of the eventual audience of their messages and contributors were informed of the origin and conditions under which appreciation was expressed. The consequence of this was that we had lower control over factors such as which packages were thanked, and what users said in their thanks messages. This also meant, we could only notify, and therefore recruit to our study, contributors who were actually thanked by users in the course of the study, regardless of the size of that population and demographic distribution.

\textbf{(2) Minimizing disruption to contributors.} Since contributors would only learn about the study through the notifications they received, we decided to deliver these privately via individual emails to avoid potential reputational effects of public notifications. To minimize disruption, we sent only one aggregated notification to each contributor with ``thanks'' across a period of time, rather than notifying them each time a ``thanks'' was received. This decision resulted in a sequencing of research activities: first, a deployment of the code editor extension with users, followed by notifying the contributors. Finally, we decided to limit the notifications to \textit{recent} contributors so as to not disturb past contributors who might have disengaged from the project. As a starting point, we decided to send notifications to only the 20 most recent contributors of the thanked software, if there were more than 20 unique contributors. Both users and contributors were made aware of this heuristic throughout the study. \ed{The decision to send the thanks after aggregation and the decision to notify the 20 most recent contributors were simple options that met our study requirements and allowed us to deploy a working probe to investigate our research questions. We do not intend to suggest that these are the best choices for a final appreciation system (see our note on implementation details in Appendix \ref{implementationnotes}).

As a result of these considerations, we decided to pursue our probe development and study activities sequentially: first, we deployed the editor extension with users, and then we aggregated the thanks messages and notified contributors.   
}

%% file: sections/study.tex
\ed{In the next section, we describe: (1) the Hug Reports code editor extension; (2) our deployment of the extension with users and research activities aimed at them; (3) our procedure of notifying contributors, and research activities aimed at them; and (4) the strategies we used to analyze the collected materials. This study was approved by our university’s Institutional Review Board.}

\subsection{The Hug Reports extension}
\begin{figure}[t]
    \centering
    \includegraphics[width=\textwidth]{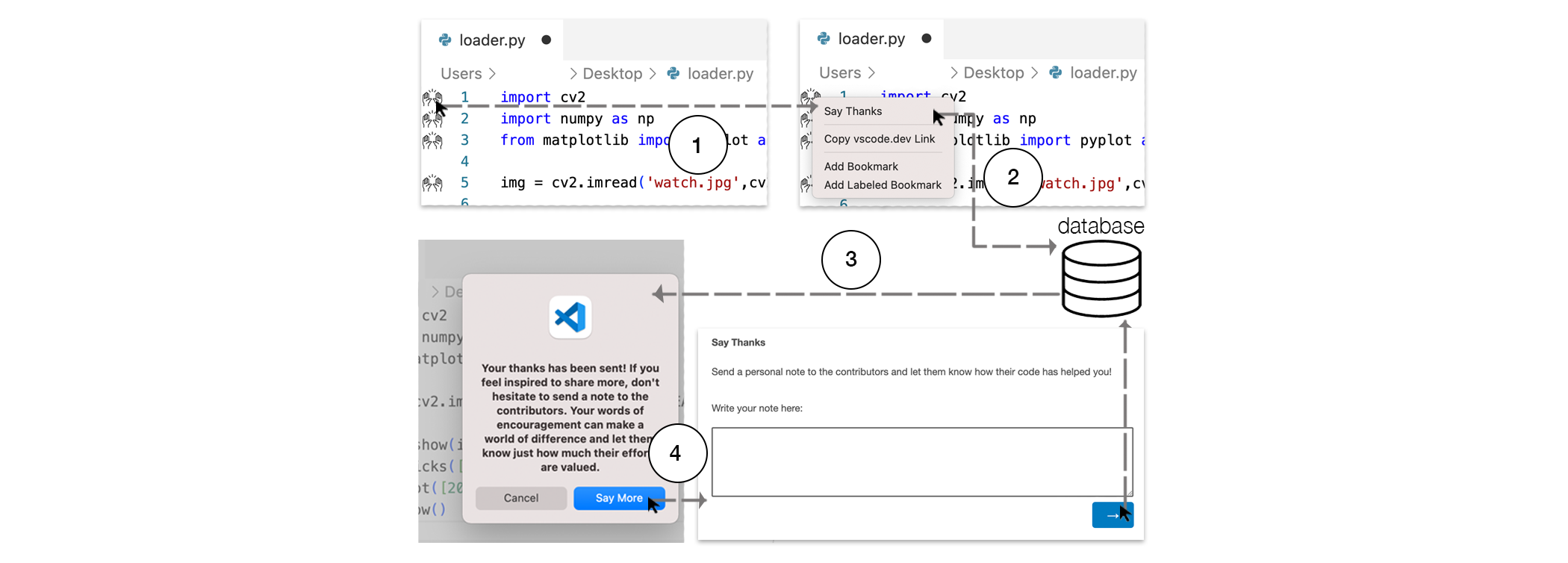}
    \caption{Hug Reports extension user flow. A button (\includegraphics[height=1em]{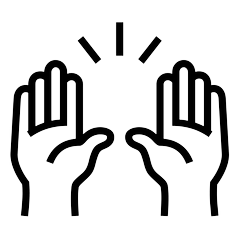}) is rendered on every line of the file that interfaces with an imported package. Right-clicking the button (1) pulls up a contextual menu with the option to ``Say Thanks''. Clicking the option (2) logs a \textit{thanks} to the cloud database, after which users are shown a success message with the option to ``Say More'' (3). If they click ``Say More'' (4), they are redirected to a web form where they can type out a \textit{personal note} that is then logged to the cloud database.}
    \label{fig:userflow}
\end{figure}
We developed an extension for the Visual Studio Code (VS Code) editor that allows users to express appreciation to developers of Python and JavaScript/TypeScript packages that they are using\footnote{JavaScript and Python are the most commonly used programming languages according to the 2023 Stack Overflow Developer Survey~\cite{DevSurvey}. Similarly, we chose to develop an extension for Visual Studio Code because it is the most popular development environment~\cite{DevSurvey}.}. It also allows them to specify whether the object of their appreciation is the entire package, specific modules, or functions within the package. The extension gets activated whenever a Python (\inlinecode{text}{.py}), JavaScript (\inlinecode{text}{.js}, \inlinecode{text}{.jsx}), or TypeScript (\inlinecode{text}{.ts}, \inlinecode{text}{.tsx}) file is opened. The user flow of the extension is described in Figure \ref{fig:userflow}. The main communication affordance of the extension is a button (\includegraphics[height=1em]{images/light.png}) that is rendered in the gutter, on every line of the file that interfaces with an imported package. \ed{The button is present next to imports of entire packages (e.g. ``\inlinecode{javascript}{import Quill from "quill";}''), imports of specific modules (e.g. ``\inlinecode{python}{from matplotlib import pyplot as plt}''), as well as lines where a function from the package or module is being used (e.g. ``\inlinecode{python}{img = cv2.imread('watch.jpg',cv2.IMREAD_GRAYSCALE)}'')}. When a user right-clicks the button on a given line, it displays a contextual menu with an option to ``Say Thanks''. Clicking that option logs a \textit{thanks} along with the line of code, which is used as a representation of the object of the user's appreciation. \ed{Clicks next to the import of a package are treated as thanks directed at the package as a whole. Similarly clicks next to imports of modules, and next to function calls, are treated as being directed at the module and function respectively.} The \textit{thanks} itself, is a one-bit signal of appreciation, similar to a ``like'' or ``upvote''. A modal pop-up notifies the user that their \textit{thanks} has been logged and gives them the option to ``Say More''. Clicking ``Say More'' redirects them to a web form in their browser where they can type out a longer \textit{personal note}. Users are not required to create an account. The \textit{thanks} as well as \textit{personal notes} do not identify their author; they are only associated with an installation ID, which is uniquely assigned for every installation of the extension. Each \textit{thanks} is logged to a database and is associated with the installation ID of the user, the line number, the line of code, and a \textit{personal note} if provided. The implementation of the extension is open source and available at: 
[anonymized for review].
Implementation notes are in Appendix \ref{implementationnotes}.

\ed{By providing a communication affordance within the code editor, and allowing users to express appreciation at the level of packages, modules, or functions rather than individual commits or pull requests, the extension lowers procedural effort in expressing appreciation. We chose to present the communication affordance via a persistent button on the interface to provide users with a subtle reminder of the contributors. 

To offer users flexibility in how much personal effort they invested, we chose the simple approach of capturing appreciation in the form of \textit{thanks} and \textit{personal notes}. This draws on the canonical approach of having a lightweight interaction (e.g. \textit{likes} and \textit{stars}) alongside an interaction for more personal effort (e.g. \textit{comments} and \textit{reviews}), which is common on other social platforms. Although we chose a simple and common approach to accomplish this flexibility, we note that other equally good options are likely available. For the purpose of the study, \textit{thanks} and \textit{personal notes} were captured anonymously. Since our research questions did not primarily concern author identities, we chose the option that led to data minimization and system simplicity.}
\subsection{Deployment with users}
\subsubsection{Participants}
\begin{table}[t]
\centering
    \footnotesize
\begin{tabular}{lrlll}
\hline
                     & \multicolumn{1}{l}{}             &                 &                                                                                                &                                                                                                                                                           \\
\textbf{Participant} & \multicolumn{1}{l}{\textbf{Age}} & \textbf{Gender} & \textbf{\begin{tabular}[c]{@{}l@{}}Programming \\ proficiency\\ (self-report)\end{tabular}} & \textbf{\begin{tabular}[c]{@{}l@{}}Weekly hours spent, \\ on average, writing \\ Python, JavaScript, \\ or TypeScript code \\ (self-report)\end{tabular}} \\
                     & \multicolumn{1}{l}{}             &                 &                                                                                                &                                                                                                                                                           \\ \hline
                     & \multicolumn{1}{l}{}             &                 &                                                                                                &                                                                                                                                                           \\
U1                   & 23                               & Man             & Advanced                                                                                       & 10                                                                                                                                                        \\
U2                   & 28                               & Man             & Advanced                                                                                       & 10-15                                                                                                                                                     \\
U3                   & 26                               & Man             & Advanced                                                                                       & 5+                                                                                                                                                        \\
U4                   & 25                               & Woman           & Intermediate                                                                                   & 15-20                                                                                                                                                     \\
U5*                   & 24                               & Woman           & Advanced                                                                                       & 40                                                                                                                                                        \\
U6                   & 28                               & Woman           & Advanced                                                                                       & 20-30                                                                                                                                                     \\
U7                   & 27                               & Man             & Advanced                                                                                       & 8                                                                                                                                                         \\
U8                   & 25                               & Man             & Intermediate                                                                                   & 25                                                                                                                                                        \\
U9                   & 26                               & Woman           & Advanced                                                                                       & 5                                                                                                                                                         \\
U10                  & 28                               & Man             & Intermediate                                                                                   & 20                                                                                                                                                        \\
U11*                  & 29                               & Man             & Advanced                                                                                       & 7                                                                                                                                                         \\
U12                  & 25                               & Man             & Expert                                                                                         & 40                                                                                                                                                        \\
U13                  & 24                               & Woman           & Advanced                                                                                       & 10                                                                                                                                                        \\
U14                  & 24                               & Man             & Advanced                                                                                       & 28                                                                                                                                                        \\
U15                  & 23                               & Man             & Advanced                                                                                       & 10                                                                                                                                                        \\
U16                  & 30                               & Man             & Intermediate                                                                                   & 45                                                                                                                                                        \\
U17*                  & 27                               & Woman           & Advanced                                                                                       & 20                                                                                                                                                         \\
U18*                  & 26                               & Man             & Advanced                                                                                       & 20                                                                                                                                                        \\
                     & \multicolumn{1}{l}{}             &                 &                                                                                                &                                                                                                                                                           \\ \hline
\end{tabular}
\caption{User demographics. The asterisk identifies users who completed the deployment requirements but did not participate in interviews.}
\label{demographics}
\end{table}
\ed{Like other design methods introduced early in the design process, field studies of technology probes focus on producing a rich qualitative account rather than statistically valid results~\cite{hutchinson2003technology, zimmerman2007research, leong2023social}. These considerations suggest a sample size that balances the researchers' abilities to do deep qualitative analysis with the ability to observe diverse participant experiences~\cite{sellen2006homenote}. Previous work in CSCW, for instance, has used a dozen dyads~\cite{leong2023social}; some prior work suggests a sample of about 10 participants~\cite{leong2023social}. To achieve a sample size in this range while accounting for the likely scenario that not all participants might complete all research activities, we halted recruitment when we had onboarded 18 participants to the study.} We recruited users through flyers distributed across the campus of a private university in the United States. Flyers were also distributed within the university community (through Slack channels and personal contacts) and publicly via Twitter.

Two criteria were used to screen potential participants. We required that they primarily use VS Code as their development environment and that they anticipated writing Python/JavaScript/TypeScript code at least 3 times a week. The screening survey also included questions about users' demographics and backgrounds. Responses are summarized in Table \ref{demographics}. The 18 users (12 men and 6 women) were aged 23-30. The survey asked users to report their experience levels as either Beginner, Advanced Beginner, Intermediate, Advanced, or Expert (categories present in the developer skill matrix). As Table \ref{demographics} shows, users described themselves as Advanced (13/28), Intermediate (4/18), and Expert (1/18). We also asked users to report their weekly average of the amount of time they spent writing code, which ranged from 5 hours to 45 hours a week. 

\subsubsection{Procedure}
We deployed the extension for three weeks so that users would have sufficient time to familiarize themselves with it and explore how they might interact with it in the course of their usual programming activities. To observe naturalistic use, required usage was deliberately kept minimal; however, we still encouraged (without enforcing) users to engage with the extension a meaningful amount so that they had opportunities to reveal their experiences, and so that there would be sufficient messages with which to understand contributor experiences. Each user was required to participate in a \ed{15-minute} onboarding session \ed{conducted over video call}, during which we helped them install the extension and provided a brief tutorial. During this session, they were also required to complete a brief pre-study questionnaire that asked them to provide: (1) an open-ended response describing how often they typically thanked contributors of packages they used; and (2) a rating on a seven-point scale (strongly disagree to strongly agree) indicating their agreement with the statement: \textit{``There are many developers whose work I am grateful for.''} \ed{These questions were intended to capture their current feelings and practices of expressing appreciation. These are presented in Table \ref{bigger-demo} (Appendix \ref{bigger-participants}) to provide more detail about the participant population}. At the end of the onboarding session, we encouraged users to send \textit{thanks} at least two times every day they found themselves coding while expressing that this would not be enforced. Users were compensated \$20 for participating in the onboarding session and completing the pre-study questionnaire. During the three weeks, for every \textit{thanks}, the extension logged the installation ID\footnote{For the purpose of the study, we established a link between participants and their installation IDs during the onboarding session.}, line of code, line number, and \textit{personal note} if added. \edit{We planned to use the \textit{thanks} logged over the course of the deployment to pursue our research activities with contributors. We also planned to analyze usage patterns to address \textbf{RQ2}.} \edit{To further investigate our research questions, }at the end of the three weeks, we invited all 18 users to participate in a one-on-one interview, 14 of whom agreed to participate (those who did not are indicated in Table \ref{demographics}). Interviews were conducted in English, remotely via Zoom, and lasted between 20 and 50 minutes, for which users were compensated \$15. Interviews were semi-structured and investigated factors that contributed to users' feelings of appreciation and decisions to express it, the experience of selecting what to \textit{thank} and what to include in \textit{personal notes}, trends in their usage, and feedback on the interface and interaction. Across these topics, we asked users if there were ways in which different or new designs could have better supported their experience. All interviews were recorded and transcribed. 

\subsection{Notifying developers}
At the end of the three-week deployment, we parsed and cleaned the extension's event data. Together, the $18$ participants logged $107$ \textit{thanks}, and $23$ \textit{thanks} included \textit{personal notes}. For the purpose of our study, we decided to map the \textit{thanks} to the package repositories and contributors manually. While we were aware of technical routes to attempt automating this\footnote{The npm and PyPI registry link to the GitHub repositories for most packages. GitHub's new search API, based on the tree-sitter and stack-graphs library, makes it possible to map a function or class name to the corresponding file (usually returning the file path as the top result). Further, GitHub's REST API provides a way to retrieve a history of all commits for a specific path, from which unique contributors can be identified. However, a fully automated solution is still technically challenging. For instance, just from an import statement in Python, it's not always possible to definitively differentiate whether the imported entity is a function, class, or submodule but such differentiation is necessary before it can be mapped to the path; the code search API is only apt for functions or classes, not submodules since it only returns files, not directories.}, we did not want to expend significant effort on developing a robust algorithmic approach for the purpose of our technology probe study. This was because the number of \textit{thanks} were few enough to be manually matched and further, we did not want to develop an algorithmic solution while other aspects of the system were still malleable (e.g. we did not want to develop a solution to detect contributors to modules or functions if the specificity was ultimately not found useful by users and contributors).  

\subsubsection{Identifying contributors:} We first identified each unique object that was thanked (the content of the line of code at which the button was clicked). For this process, we treated packages, modules, and functions as separate objects since the 20 most recent contributors to each could be different. \ed{Thanks logged next to the import of a package were treated as thanks directed at the package as a whole. Similarly, thanks logged next to imports of modules, and next to function calls, are treated as being directed at the module and function respectively.} We note that each object could have been thanked multiple times (e.g. ``\inlinecode{python}{import cv2}'' was thanked twice). Corresponding to each unique object, we maintained a count of the number of \textit{thanks} it was associated with and all \textit{personal notes} associated with that object. We identified a total of $70$ unique objects that were thanked, most just once. We then mapped each object of appreciation to its source code on GitHub. \ed{Thanked packages were mapped to their repositories, while thanked modules and functions were mapped to their corresponding file. We used this to then find the 20 most recent contributors. For thanked packages, we identified the 20 most recent contributors to the entire repository. For thanked modules and functions, we identified the 20 most recent contributors to the corresponding file.} Every commit, contains an email address of the author of the code patch, which provided us with contact information for the contributors. \ed{In determining the 20 most recent contributors, we skipped commits where the provided email address was anonymized (emails ending in ``users.noreply.github.com'')}. If the total number of unique contributors was fewer than 20, we recorded all contributors. This process resulted in a total of $550$ contributors to be notified. $470$ had been thanked for $1$ object each, with the remaining being thanked for $2$ to $8$ objects each.  

\subsubsection{Notification}
\begin{figure}[t]
    \centering
    \includegraphics[width=\textwidth]{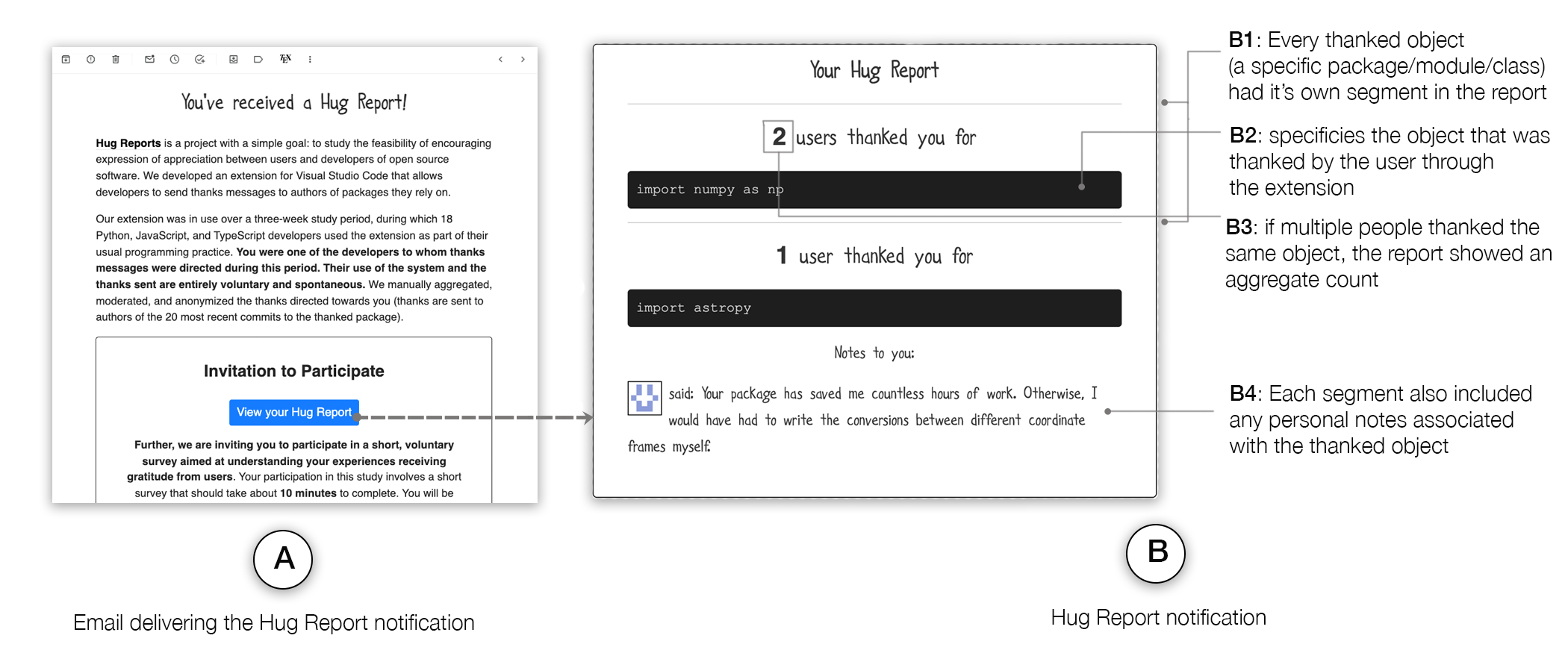}
    \caption{(A) The Hug Report notification was delivered via an email that provided the contributors with context about our project so that they could understand the conditions under which the users had directed these \textit{thanks} and \textit{personal notes} towards them. (B) The Hug Report notification, itself, was organized so that each segment (B1) showed one object that the contributor had been thanked for (B2). Above the object, the notification showed the number of \textit{thanks} corresponding to it (B3). Below the object, were any \textit{personal notes} associated with it (B4)}
    \label{fig:notification}
\end{figure}

As shown in Figure \ref{fig:notification} (B), the notification was organized so that each segment showed one object that the contributor had been thanked for (B1). Above the object, the notification showed the number of \textit{thanks} corresponding to it (B3). Below the object, we included any \textit{personal notes} associated with it (B4). The notification was delivered via an email (as shown in Figure \ref{fig:notification} (A)) that provided the contributors with context about our project so that they could understand the conditions under which the users had directed these \textit{thanks} and \textit{personal notes} towards them. \ed{The design of the notification message was modeled after notifications in Wikipedia's ``Thanks'' system~\cite{matiasdiffusion}.}

\subsubsection{Procedure and Participants}
To understand contributors' reactions to the notifications, the email also included an invitation to participate in our study with a link to a survey. The survey included questions on demographics including age, gender, programming proficiency, and their tenure in the project for which they were thanked. It then asked contributors how they felt on three \ed{single-item} scales adapted from prior work~\cite{kumar2018undervaluing}: (1) a scale ranging from -5 (much more negative than normal) to 5 (much more positive than normal), with the midpoint of 0 labeled no different than normal; (2) a scale ranging from -5 (not at all surprised) to 5 (extremely surprised); and (3) a scale ranging from -5 (not at all awkward) to 5 (extremely awkward). We also included an open-ended question asking contributors to further describe how the notification made them feel. The survey included three more open-ended questions: (1) \textit{Prior to this, how often did you receive messages of thanks from users?}; (2) \textit{What else would you have liked to know about the senders of these thanks?}; and (3) \textit{Do you have any feedback for us that you would like to share?} Finally, the survey invited contributors to participate in an interview. There was no compensation for participating in the survey and all questions on the survey were indicated as optional.

We received 26 responses to the survey (4.7\% response rate). This included 23 men and 3 women. Respondents were aged 21-56 (median age was 34). 13 respondents had been involved for less than a year in the project for which they were thanked. Five respondents had been involved for 1-3 years, five had been involved for 3-5 years, and two respondents had been involved in the thanked project for more than 5 years (1 respondent did not disclose their tenure). In describing our findings, contributors who responded to the survey are labeled as C\# (C1 to C26). 

\begin{table}[t]
\centering
    \footnotesize
\begin{tabular}{lrlll}
\hline
                     & \multicolumn{1}{l}{}                                       &                                                            &                                                                                                &                                                                                                    \\
\textbf{Participant} & \multicolumn{1}{l}{\textbf{Age}}                           & \textbf{Gender}                                            & \textbf{\begin{tabular}[c]{@{}l@{}}Programming \\ proficiency\\ (self-report)\end{tabular}} & \textbf{\begin{tabular}[c]{@{}l@{}}Tenure in project\\ for which they\\ were thanked\end{tabular}} \\
                     & \multicolumn{1}{l}{}                                       &                                                            &                                                                                                &                                                                                                    \\ \hline
                     & \multicolumn{1}{l}{}                                       &                                                            &                                                                                                &                                                                                                    \\
C5                   & 23                                                         & Man                                                        & Intermediate                                                                                   & \textless{}1 year                                                                                  \\
C8                   & \begin{tabular}[c]{@{}r@{}}did not disclose\end{tabular} & Man                                                        & Expert                                                                                         & 1-3 years                                                                                          \\
C12                  & 45                                                         & Man                                                        & Advanced                                                                                       & \textless{}1 year                                                                                  \\
C13                  & 36                                                         & Man                                                        & Advanced                                                                                       & \textless{}1 year                                                                                  \\
C18                  & 24                                                         & Woman                                                      & Advanced                                                                                       & 1-3 years                                                                                          \\
C19                  & 30                                                         & Man                                                        & Expert                                                                                         & \textless{}1 year                                                                                  \\
C20                  & 32                                                         & Man                                                        & Advanced                                                                                       & \textless{}1 year                                                                                  \\
C21                  & \begin{tabular}[c]{@{}r@{}}did not disclose\end{tabular} & Man                                                        & Advanced                                                                                       & \textless{}1 year                                                                                  \\
C23                  & \begin{tabular}[c]{@{}r@{}}did not disclose\end{tabular} & Man                                                        & Expert                                                                                         & \textless{}1 year                                                                                  \\
C27                  & \begin{tabular}[c]{@{}r@{}}did not disclose\end{tabular} & \begin{tabular}[c]{@{}l@{}}did not disclose\end{tabular} & Expert                                                                                         & \textless{}1 year                                                                                  \\
                     & \multicolumn{1}{l}{}                                       &                                                            &                                                                                                &                                                                                                    \\ \hline
\end{tabular}
\caption{Contributors who participated in the interviews.}
\label{contributor-demographics}
\end{table}

Of the respondents, $10$ agreed to be interviewed. Interviews were conducted in English, remotely via Zoom, and lasted 30 minutes. Interviews were semi-structured and focused on the experience of receiving \textit{thanks}, scoping of appreciation, approaches for attribution, frequency and form of notifications, and their ideas for new or different design concepts. Contributors who participated in the interview were compensated \$15. One contributor, labeled C27, did not respond to the survey questions but agreed to be interviewed. Table \ref{contributor-demographics} shows demographic information of contributors who participated in interviews.

\subsection{Analysis}
\ed{We used the usage data collected from the extension deployment as one source of data to address \textbf{RQ2}. From this, we created summary statistics of the interactions that took place. We also used an affinity diagramming approach to group personal notes into categories, based on their content. We used contributors' responses to the survey, as one source of data to address \textbf{RQ1} and \textbf{RQ2}. We created summary statistics for responses to quantitative questions and used an affinity diagramming approach to analyze the responses to open-ended questions. In this way, we used these two sources of data to derive descriptive summaries of the behaviors of users and reactions of contributors.

We used the interviews to address \textbf{RQ1}, \textbf{RQ2}, and \textbf{RQ3}.} To do this, we conducted a reflexive thematic analysis~\cite{braun2006using}. Two of the authors independently performed a line-by-line open coding of transcripts from the first 7 (out of the 14) user interviews and the first 5 (out of the 10) contributor interviews. Codes generated in this phase were in part inductive, driven by the data, and in part guided by our original research questions– we remained open to capturing observations that emerged through the data while also looking out for observations that related to our main guiding questions. All authors met to discuss the analysis and iteratively refined the codes, following which the first author applied the refined set of codes to the remaining transcripts looking at whether participant experiences fit into our existing categories. Finally, all authors discussed the analysis to iteratively refine and solidify the themes, and group similar themes together. \ed{Themes were generated at a semantic level,
reflecting what participants explicitly said~\cite{braun2006using}.} 

%% file: sections/findings.tex
\ed{In this section, we begin with findings from the descriptive analysis of usage data and contributors' responses to the questionnaire and survey (\ref{overview}). \edit{In it, we provide a descriptive summary of the extension's use and contributors' reactions, addressing \textbf{RQ1}, and \textbf{RQ2}. Across the next subsections, we present the main themes from our analysis of the interviews. In section \ref{affordances}, we first discuss the extent to which key decisions of the extension encouraged meaningful appreciation (\textbf{RQ1}). In sections \ref{moments}, \ref{meanings}, and \ref{attribution} we present findings that speak to \textbf{RQ2}, describing how appreciation was felt, expressed, received, and interpreted. Section \ref{moments} describes the moments in which users expressed appreciation. Section \ref{meanings} describes the two meanings that appreciation took on: (1) as a measure of utility, where the volume of \textit{thanks}, and what it was directed at, were interpreted as a signal of the software's utility, and (2) as an act of expressive communication that intended to convey a user's gratitude. Users' interactions and contributors' interpretations depended on which of the two meanings they prioritized in a given context (\ref{meanings}). In section \ref{attribution}, we discuss how contributors' reactions were influenced by how much they felt they had contributed to the object that was being appreciated. Finally, we present participants' ideas (\ref{feedback}) for further supporting exchanges of appreciation, addressing \textbf{RQ3}.} } 
\subsection{Descriptive analysis}
\label{overview}
\begin{figure}[t]
    \centering
    \includegraphics[width=\textwidth]{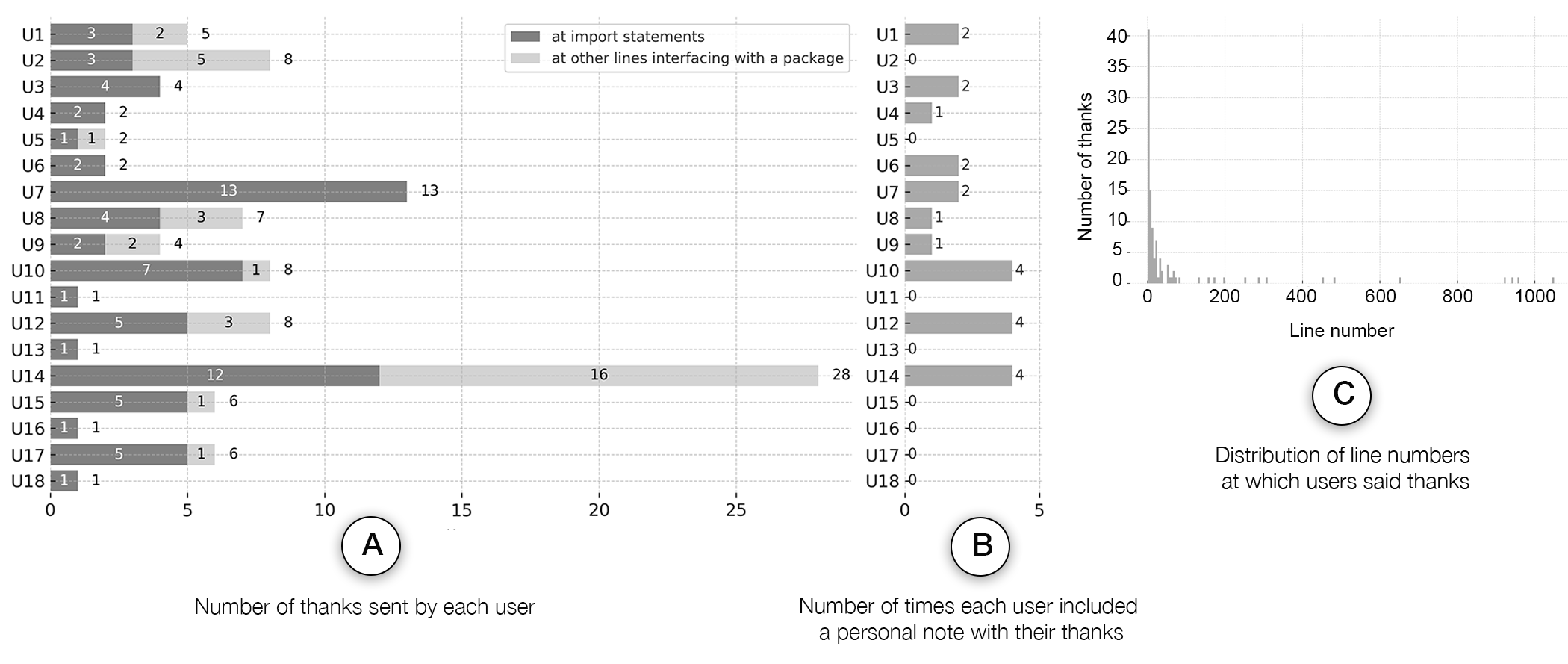}
    \caption{Summary of users' engagement with the Hug Reports extension. (A) The 18 users logged a total of 107 \textit{thanks}. \textit{Thanks} were expressed more often at import statements (72 instances) than other lines interfacing with a package (35 instances). (B) Users included \textit{personal notes} with their \textit{thanks} on 23 occasions. (C) The affordance was more often accessed near the start of a file than later on.}
    \label{fig:usage-stats}
\end{figure}

\subsubsection{Patterns of use \ed{(RQ2)}} 
\label{usepatterns}
Over the course of the three-week deployment, the $18$ participants logged $107$ \textit{thanks}, and $23$ \textit{thanks} included \textit{personal notes} There were considerable differences in the frequency with which different users interacted with the extension (see Figure \ref{fig:usage-stats} (A) and (B)). We found that the lines of code at which the affordance was clicked, we more often import statements (72 instances) than other lines interfacing with a package (35 instances) (see Figure \ref{fig:usage-stats} (B)). Relatedly, the affordance was more often accessed near the start of a file than later on (Figure \ref{fig:usage-stats} (C)).

\textit{Personal notes} varied in their content. Some (7/23) stopped at expressing general sentiments of gratitude towards the developers (e.g. \textit{``Thanks! I'm super relying on this!''}, expressed at: ``\inlinecode{python}{import pandas as pd}''). Some notes (6/23) additionally described how they appreciated the general functionality the package was intended to provide (e.g. \textit{``Numba made my pandas code so much faster!''}, expressed at: ``\inlinecode{python}{from numba import njit}''). 5 of the 23 notes, appreciated specific design choices made by contributors, such as supporting interoperability with other utilities, helpful error messages, and effective organization/modularization  (e.g. \textit{``Thanks for making this library work so well with numpy so I don't have to write much extra code to support sparse matrices!''}, expressed at: ``\inlinecode{python}{from scipy import sparse as sps}''). Three of the notes mentioned the ways in which users were personally using the packages in their work (e.g. \textit{``This has been so useful for our project. We're trying to do satellite localization using a camera and we're essentially just fine-tuning your model on our dataset and wrapping it around a state estimator. It works so well!''}, expressed at: ``\inlinecode{python}{from ultralytics import YOLO}''). In two notes, users revealed aspects of their background that shaped their appreciation (e.g. \textit{``I'm not a machine learning/computer vision scientist. OpenCV has helped me a lot with video processing!''}, expressed at: ``\inlinecode{python}{import cv2}''). Finally, two notes expressed general sentiments of appreciation towards the organization working on the package (e.g. \textit{``Thanks folks at Microsoft for making TypeScript''}, expressed at: ``\inlinecode{javascript}{import * as ts from 'typescript';}''). 

\begin{figure}[t]
    \centering
    \includegraphics[width=\textwidth]{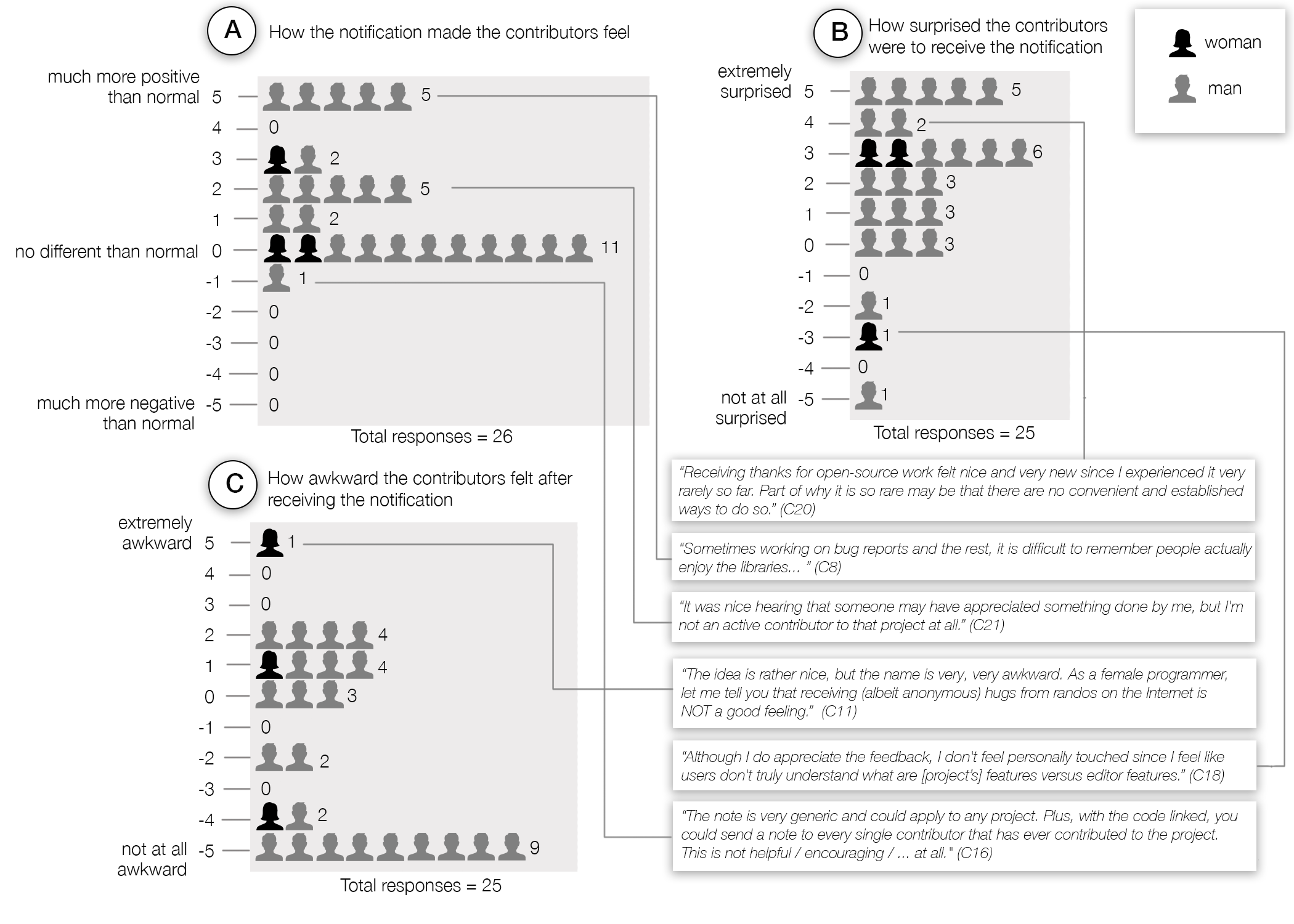}
    \caption{Summary of contributors' responses to the survey. We include some contributors' responses to the open-ended questions in the survey to provide further context for some responses. (A) Most contributors reacted positively to the notification, and only 1 participant reacted negatively (whose survey response suggests it was because they found the \textit{thanks} and \textit{personal note} too generic). (B) Most contributors were surprised to receive the notification, however, few contributors who belonged to large projects were not. (C) Most contributors felt no or low levels of awkwardness being appreciated. One contributor, who felt `extremely awkward' noted that receiving a `hug' from a stranger felt awkward. }
    \label{fig:contributer-stats}
\end{figure}

\subsubsection{Contributors' responses \ed{(RQ1, RQ2)}}
\label{contributor-responses}
Overall, contributors reacted positively to receiving appreciation. Contributors hoped to know more about why they were being thanked and would have liked more personalized messages. Some contributors didn't feel like they deserved the appreciation if they hadn't substantially contributed to the package for which they were thanked. Figure \ref{fig:contributer-stats} summarizes the responses we received to the survey, from 26 contributors. 14 contributors reported that receiving the notification made them feel more positive than normal (Figure \ref{fig:contributer-stats} (A)), 11 reported feeling no different than normal, and only 1 participant reacted negatively (whose survey response suggests it was because they found the \textit{thanks} and \textit{personal note} too generic). Responding to how often they currently received appreciation, most contributors reported it was rare, some suggesting they received \textit{``a few per year''} (C15), such as in \textit{``GitHub Issues once every couple months''} (C13) or \textit{``once a feature request is closed, primarily with emojis, and it is extremely rare to get a text message''} (C23). Participants further described how the notification made them feel. One mentioned how \textit{``receiving the hug report was a wonderful surprise, and brightened [their] day''} (C3), while another mentioned they \textit{``felt happy and feel motivated to do more''} (C1). One contributor explicitly mentioned how current channels for appreciation are few: \textit{``receiving thanks for open-source work felt nice and very new since I experienced it very rarely so far. Part of why it is so rare may be that there are no convenient and established ways to do so''} (C20). However, few contributors noted how there was \textit{``nothing personalized about the report''} (C9). Without a more specific message, C18 mentioned: \textit{``although I do appreciate the feedback, I don't feel personally touched''}. Contributors described wanting to know more about why they were being thanked and what users were using their packages for (e.g. \textit{``more than the one sentence of thanks, what was useful for them that they were thanking for?''} (U19) and \textit{``having a rough idea about types of projects my features are used for is quite beneficial.''} (C23)). Most contributors were surprised to receive the notification (Figure \ref{fig:contributer-stats} (B)), however, few contributors who belonged to large projects were not. Most contributors felt no or low levels of awkwardness being appreciated (Figure \ref{fig:contributer-stats} (C)). Some participants felt awkward because they did not feel they had contributed significantly to the project for which they were being thanked (we discuss this further in \ref{attribution}). One contributor, who felt ``extremely awkward'' noted that a `hug' from a stranger felt awkward (quote in Figure \ref{fig:contributer-stats}). 

\subsection{Hug Reports encouraged appreciation that was still meaningful to users and contributors \ed{(RQ1)}}
\label{affordances}

\subsubsection{The button served as an ambient reminder of contributors' effort}Many users felt the button in the gutter made them grateful more often, which was described as a welcome feeling:
\begin{quote}
``I think a lot of times I'm kind of just like in the weeds programming and you kind of forget that the code that you're using is someone else's code ... with that like task oriented mindset, it can be hard to pop up a level and actually appreciate the effort that someone else went through to produce this code. The humanizing factor, I think, is critical to make like necessity of expressing gratitude like real. Even if I'm not actually pressing the thanks button all the time, having the reminder that these are the packages I'm using and there are real people that made them ... even having that mindset as a byproduct of having the extension installed is nice.'' (U6)
\end{quote}

U15 described how his sense of gratitude extended beyond the packaged code that the extension supported:
\begin{quote}
``That was actually something interesting that the tool made me mindful of...that there are contributors to my development experience that I wasn't necessarily thinking of..like what else is going on in the environment I use. Oh, ZSH! I never think to thank ZSH! Or the maintainer of Brew.'' (U15)
\end{quote}
Users suggested the reminder supported their own intentions to be more appreciative (e.g. \textit{``I wish I naturally thought about that more''} (U6); \textit{``I appreciated it supporting the thing that I believe in''} (U9); and \textit{``I always want to be as gracious, or, you know, as grateful as the situation allows''} (U15)).

\subsubsection{Low procedural effort encouraged appreciation}Users also described how the extension made it easier to express appreciation:\textit{``it's pretty straightforward, just like one click away''} (U8). Users felt the extension made expressing appreciation more approachable than current channels:
\begin{quote}
``It felt like a nice opportunity for me to thank them without having to find out how to contact them. Right now, you don't have any way of sending thanks and so, if you lower the barrier of sending thanks compared to `I have to find where to GitHub repo is, and leave an issue', I would assume it would increase the number of people who actually send thanks.'' (U2)
\end{quote}

Further, users liked having the option to express appreciation as a \textit{thanks} since it was low effort enough that it didn't disrupt their flow and it gave them a way to express appreciation when they didn't \textit{``know what to say''} (U9). Contributors acknowledged this sentiment behind the \textit{thanks}: \textit{``I think it's helpful. Because sometimes you don't know what to say in a comment. You don't have something particularly meaningful, but you still appreciate it''} (C12). 14 of the contributors who responded to the survey had received only a \textit{thanks} without any \textit{personal notes}, 5 of whom reported feeling no different than normal, 3 reported feeling much more positive than normal, and the remaining 6 reported positive emotions in between the two. None of them reported feeling negatively. Since the extension gave users flexibility in how much effort they could convey, C13 suggested: \textit{``in this hierarchy of showing support you go all the way from not doing anything to supporting monetarily right? And somewhere along the rung, there is this, the starring of the GitHub Repo. And at a slightly higher, more personalized level, there is \textit{thanking}''}.   

\ed{\subsection{Moments in which users expressed appreciation (RQ2)}\label{moments}}
\subsubsection{Users engaged with the extension when they ``came up for air''} Users described how they sent \textit{thanks} in moments of transition between tasks, or moments of \textit{``rest''} (U7) between tasks. These moments often coincided with transitions between files and so, import statements at the top, where the file would open, were often the site where users sent \textit{thanks}. U6 described these as moments when she was \textit{``coming up for air''}, elaborating:
\begin{quote}
I think [sending \textit{thanks}] is something that I did when I wasn't really in the middle of any sort of programming task. Having them at the top, like with all the imports, is kind of nice, because it's naturally, where you kind of start within a file. So, if I open a file I'm like, `oh, yeah there's all these imports like I should thank them'. It's like, I'm already kind of in a paused state. I'm getting ready to do something, or I just finish something.
\end{quote}
When users were focused on a task, \textit{``the icon became sort of like a background noise''} (U1). U2 and U14 shared a similar reflection: \textit{``When I'm down in a file I really wanna focus on writing the code rather than stopping and saying thanks''} (U2) and \textit{``I didn't really want to break my coding flow, so I would usually club them once I had some part of some module running''} (U14). 

\subsubsection{Users thanked packages retrospectively, repeating if they discovered new use cases}
Even though thanks was expressed at the top of the file, and in transitional moments, it was not expressed preemptively. Users mentioned how they felt appreciative, and thanked packages, once they had got them to work for their specific use case: \textit{``It's like it did the thing. It has shown me that it can do the thing. And this is a good thing''}(U9). As U15 put it: \textit{``I wasn't saying good idea, but good implementation''}. Similarly, U3 mentioned: \textit{``After I ran the code, then I would scroll up and say thank you.''}. U14 also described scrolling up retrospectively: \textit{``whenever I saw that I had completed one module, I would just go at the top and see what modules I've used''}. Further, users often described feeling appreciative when they discovered new functionalities. U7 mentioned: \textit{``I'd like to say thanks if I find something useful or something I didn't know before. I think saying thanks sort of strengthens my memory of my experience with that specific function.''}. Users also mentioned how they would be inclined to thank packages again if they found new use cases, suggesting that \textit{thanks} can be meaningfully different than one-time exchanges like starring on GitHub. U9 suggested: \textit{``I think if there is a unique functionality that hadn’t been captured before, so like you know, I have use case A, and then I find out there’s a use case B, I would definitely thank again.''}

\subsubsection{Broader rhythms in work influenced appreciation} Users described how their engagement with the extension was also influenced by broader rhythms in their work, where they found themselves more prone to reflection during certain periods: \textit{``I have my Tuesday meetings. I would just link this to my meeting, and I was like after the meeting. I'll do this. And so I would do this like bout of thanksgiving(?)''} (U10). U1 described how project cycles could also prompt reflection: \textit{``if I'm at the end of a project cycle, then I might feel gratitude again''.} U3 drew an analogy to borrowing a tool: \textit{``If someone gave you a tool for you to use, I feel like you would say thank you to them once when you get the tool and a second time when you return the tool.'' }

\ed{\subsection{Two meanings of appreciation \ed{(RQ2)}}
\label{meanings}
Users and contributors interpreted appreciation as both: (1) a \textit{measure of utility}, where the volume of \textit{thanks}, and what it was directed at, were interpreted as a signal of the software's utility, and (2) as an \textit{act of expressive communication} that intended to convey a user's gratitude. Users' interactions and contributors' interpretations depended on which of the two meanings they prioritized in a given context.

\subsubsection{Appreciation as a measure of utility}
Despite our small-scale deployment, users and contributors saw value in the quantitative metrics that could be derived from the \textit{thanks}. Some contributors envisioned metrics as being useful for providing external evidence of their own impact (e.g. \textit{``I wish this was integrated with GitHub or LinkedIn badges. It would be good and motivational to share it with my professional network''} (C1)) as well as the project's impact: \textit{``We would find it useful because funding agencies want to know that...It's only the funding people who are like `But are people using this?'''} (C8).
To access its informational value, contributors preferred a publicly shareable aggregation of data (C19, C13). C13 suggested how the system could \textit{``keep aggregating the stats on a website''}. 

When considering this interpretation of the \textit{thanks}, some contributors found value in letting users thank specific modules or functions. Few contributors also noted how this could provide a deeper understanding of the software's utility and could also be valuable for decision-making in a project:
\begin{quote}
``That's a very important metric for a maintainer, because they want to know which parts of the project are well used, which parts are not that properly used, and it can decide a lot of the trajectory of the project going forward.'' (C21)
\end{quote}
C27 described how this could provide a more representative statistic of use than currently available metrics:
\begin{quote}
``In npm you've got those traffic stats like this got downloaded 500,000 times right. But the reason a lot of packages get downloaded very often is because they are part of this massive npm package that has 700 dependencies and you need to install all of them to just use like 4 lines of code from the top-level directory but you never actually hit that line of code. So you don't end up using it, so to speak. this could be a slightly more human-centric way of understanding that'' (C27)
\end{quote}

Users, when prioritizing this interpretation of \textit{thanks}, saw lesser value in investing personal effort, and focused more on providing contributors with an informative signal:
\begin{quote}
``I was definitely more biased to [sending \textit{thanks}] on something versus \textit{personal notes}. It doesn't matter how many people say your code is the best, in terms of being able to use it for performance reviews or being able to prove its value. Some people are more biased to numeric metrics, like how many people are watching your repository, or how many people star your repository. Like okay, this will make them feel warm and fuzzy but I really just want to give them something that they can use to prove that their code is valuable to an external audience.'' (U9)
\end{quote}

\subsubsection{Appreciation as an act of expressive communication}
Users and contributors also saw appreciation as a personal communication of gratitude. When prioritizing this interpretation of appreciation, contributors expressed how they would have liked to receive more personal effort:
\begin{quote}
``I much prefer hearing when a piece of code made a true difference to someone, e.g. if they wrote `this function in [project] shaved two months off of my PhD' or `this library you wrote completely transformed the way we were able to execute our project' or even `I love the API you've designed so much!''' (C9)
\end{quote}
Other contributors also described how they would have liked to hear more specific messages, about what the users were using the package for, and what the users appreciated. 

While some \textit{personal notes} did mention these aspects (\ref{usepatterns}), many users struggled to find something specific to say, in the moment, which led to them writing generic notes of appreciation or prevented them from writing \textit{personal notes} altogether. U7 mentioned he was uncertain about what would make \textit{``good communication in this relationship between the user and contributor''}. Other users mentioned how it was a \textit{``bit tricky to explain why, I'm actually grateful''} (U12) and how they sometimes \textit{``didn't have anything specific to say''} (U6). U9 suggested: \textit{``it was difficult to know what to say other than thanks. It's a lot easier to write something when things go wrong.''}

When considering \textit{thanks} as a personal expression of gratitude, some users described how their experiences were affected by the size of the anticipated audience of their \textit{thanks} (U6, U12). In the context of larger packages, these users felt that \textit{thanks} scoped to the entire package were less meaningful than \textit{thanks} scoped to individual modules or functions.
}
\ed{\subsection{Contributors' reactions were influenced by how much they felt they had contributed to the object that was thanked (RQ2)} \label{attribution}}
Contributors felt that \textit{thanks} was undeserved if felt they had not substantially contributed to the thanked code. Even if contributors had made code edits to the part of the package that was thanked, they felt it was undeserved if they did not feel a sense of ownership. For instance, in their survey response, C20 described:
\begin{quote}
``The thanks I received were for a Python package that I happily used myself and contributed a minimum amount to (two lines of code around two years ago or such). Therefore, I don't believe I deserve praise for my contributions. However, delivering and receiving thanks gives me a good feeling, as I am happy for the actual contributors to be acknowledged.'' (C20)
\end{quote}
Other contributors had similar reactions: \textit{``I've only submitted a few patches to [project]''}(C24) and \textit{``It was nice hearing that someone may have appreciated something done by me, but I'm not an active contributor to that project''} (C21). C21 further elaborated:
\begin{quote}
``It depends on my [sense of] ownership of that project. If that is above a certain threshold, then that thanks is directly meaningful to me. So something like 5\% or 10\% would be that threshold that comes to my mind. If I've written more than 10\% of the code in the project, I would want to know when someone is thanking the project.''(C21)
\end{quote}

This is not necessarily a critical failure of Hug Reports. Some contributors felt that these \textit{``misfires''} were not an issue: \textit{``spreading the thanks signal out into the world is an example of something that, even if it misfires, it doesn't hurt and so, I don't think that the misfires are necessarily bad''} (C13). C18 described how she \textit{``wouldn't feel weird''} if thanks were misdirected at her, further adding: \textit{``it would be really easy for me to redirect it to the right person''}. Still, we recognize that better-designed heuristics for identifying relevant contributors can enhance the meaningfulness of the \textit{thanks}. 

\ed{\subsection{Ideas from users and contributors (RQ3)}
\label{feedback}
\subsubsection{Nudges} Users proposed ideas of how the extension could encourage appreciation further. For instance, U3 suggested having an ambient indicator of usage: \textit{``the button could glow or become a bit more colorful if you have been programming with that package for a while, especially, you know, if you have that package imported that you use a lot of different components from it.''} U1 suggested a similar idea: \textit{``simple threshold based things, after you've imported something 10 times, you
show like a little popup or in the status bar''}. U2 mentioned how such reminders could make usage more apparent: \textit{``These are not things that I realize until I see the statistics. So, it would be great to have some feature that reminds me of that.''} Further, U6 mentioned how the extension could also identify which packages had smaller teams, as those would feel more meaningful to thank.

\subsubsection{Opportune moments}
Reflecting on patterns in when they tend to express appreciation, users pointed out ways in which the system could harness opportune moments. U1 suggested: \textit{``every time you open or close a file maybe you could get a pop up which says that, `oh, in this file you use these things'''}. U7 had a similar suggestion, mentioning how he would like such notifications to emphasize new features he used that week. U9 and U10 mentioned how they would like to view statistics of their use, at times that they could control (e.g. immediately after a specific recurring meeting), either via a notification or dashboard. C12 suggested that if users were themselves publishing work to GitHub, the end of project cycles could be opportune moments to remind users of the packages they were using: \textit{``GitHub, has a feature where you say like, make a release right? You could tie an action into that where it would say, `Oh, you're making a release. Here are all the [packages] you used in your project in this release''' }(C12).

\subsubsection{Form and frequency of notifications} Contributors described how the form in which Hug Reports were aggregated and presented could be broadened to support different use cases:  \textit{``the psychological benefits of it would be garnered more by sending the emails out, and the economic benefits might be garnered better by putting it up on a website and sharing the link around''} (C13). Decision-making and fund-raising activities could benefit from more aggregated statistics: \textit{``from the perspective of the project, we wouldn't really care about the personal notes, because we would aggregate over them anyway''} (C8). Contributors also suggested venues to make the aggregate data public, such as in project \textit{``release notes''} (C8) and in individuals' \textit{``GitHub profiles''} (C27). C20 and C18 also felt having a way to display on the project's GitHub page, for example as \textit{``a GitHub badge''} (C18) would be nice: \textit{``on GitHub, I look at the number of stars, the number of different issues and so, it seems the number of happy users, would also be a good metric''} (C18)

\subsubsection{Scaffolding for personal effort} Both users and contributors suggested it would be useful to have prompts that could help users write more specific \textit{personal notes}: \textit{``maybe giving some instruction of like to users like, here's an example. What you can write like here was my used case, and why it was useful.''} (U9). C19 suggested having something similar to product reviews that ask users \textit{``what's good about it? Is the make good? Is the design good?''}. U3 suggested it could be useful to be able to \textit{``send pictures, you know, especially if things are hardware related''}. In his survey response, C26 suggested having discrete pre-determined categories: \textit{``maybe a selection from a few pre-determined thank-types. e.g. `I am using it for everything, thanks!' or `Saved me a ton of time in my current project'''}.

\subsubsection{Heuristics for notifying contributors}
Contributors suggested that accuracy in notifying relevant contributors was less critical since misdirected \textit{thanks} were acceptable. In fact, many contributors suggested relaxing the criteria further. Some suggestions included sending \textit{``an automated mail on the project mailing list''} (C8) or notifying \textit{``all the people who have ever committed code''} (C13). C20 suggested:\textit{``people might just thank popular functionality of a package that you did not contribute anything to but
it can just go out to everyone...that seems like the most general approach.''}

}

\edit{
\subsection{Summary of findings}
Addressing \textbf{RQ1}, our deployment found that by lowering the procedural effort and by reminding users of the contributors, Hug Reports encouraged appreciation in ways that were meaningful to users and contributors (\ref{affordances}, \ref{contributor-responses}). Addressing \textbf{RQ2}, we described trends in usage (\ref{usepatterns}), and contributor's reactions (\ref{contributor-responses}). Further addressing \textbf{RQ2}, the interviews revealed patterns in \textit{when} users expressed appreciation (\ref{moments}), how appreciation took on two meanings (\ref{meanings}), and how contributors' reactions were influenced by their perceived level of involvement (\ref{attribution}). Finally, addressing \textbf{RQ3}, we presented ideas proposed by users and contributors for how appreciation could be better supported (\ref{feedback}).}

%% file: sections/discussion.tex
\ed{Here, we begin by discussing how our work contributes new knowledge on designing appreciation systems in peer production by exploring the implications of re-orienting appreciation around abstractions of use. Then, we discuss how our work extends literature on appreciation in open source by uncovering how appreciation is experienced and expressed by users in practice. Finally, we discuss limitations of our work and directions for future work.

\subsection{Opportunity and limitation of re-orienting appreciation around abstractions of use}
Our approach tried to enable users to express appreciation in terms of the abstractions they are exposed to---the package or its modules---rather than lower-level units of contribution such as individual commits or pull requests. This made it unique from Wikipedia's ``Thanks'' system which captures appreciation towards individual edits. Our findings show how re-orienting appreciation around packages and modules meant it aligned more with how users experience appreciation and made it easier for them to express it. At the same time, shifting appreciation away from low-level units of contribution limited the extent to which appreciation could serve as a form of individual recognition. Here, we discuss the opportunity and limitations of this approach.
\subsubsection{Opportunity: Encouraging collective recognition and improving visibility into usage} First, our work suggests that capturing appreciation in terms of abstractions of use has the potential to encourage more interactions. Even if it is difficult to derive \textit{individual} recognition from such appreciation, the fact that contributors reacted positively, suggests it could still provide a valuable form of \textit{collective} recognition. Further, it gives members of the project new insights into what the project's users appreciate. Many contributors in our study pointed out the value of \textit{thanks} as a source of information that could also support project-level decision-making. Taken together, we suggest that this approach could be a valuable supplement to existing appreciation systems that capture appreciation towards low-level units of work. Following the recommendations of contributors in our study, it could be valuable to aggregate such appreciation and make it available to members of the project, even if the system does not solve the challenge of individual attribution.

\subsubsection{Limitation: Deriving individual recognition is challenging} Even though contributors reacted positively to the appreciation they received, the extent to which they thought they had `claim' to the appreciation depended on how much they felt they had contributed to the work. This foregrounds an important question: To what extent can systems derive individual recognition after capturing appreciation around abstractions of use? 

Prior work suggests that there is a limit to what automated approaches can accomplish. This is because it is challenging to objectively determine a contributor's relative impact based solely on visible contribution activities. Visible activity traces may not reflect contributions such as intellectual contributions~\cite{howison2013incentives}, code review~\cite{young2021contributions}, governance~\cite{young2021contributions}, and fund raising~\cite{young2021contributions}. Several efforts have emerged to address this limitation. For instance, the All Contributors\footnote{https://allcontributors.org/} project aims to standardize credit files and bring visibility to non-code contributions. Gitmoji\footnote{https://gitmoji.dev/about} aims to provide a standardized way to annotate commits based on the kinds of contribution they are aiming to make. Additionally, several researchers have recommended that team member roles be explicitly recorded~\cite{alliez2019attributing, casari2021open, ramin2020more}. But until these approaches are adopted as a standard, there is unlikely to be a generalized approach to deriving individual recognition. Systems attempting to do so would have to work closely with individual projects and either: (1) share the messages of appreciation with the project that members of the project can then redirect to appropriate contributors, or (2) develop heuristics that are fine-tuned to each project based on its specific practices of recognizing contributors (e.g. NumPy lists contributors in its release notes). 

Finally, even if these approaches can more accurately ascertain a contributor's relative impact in objective terms, contributors may still feel uncomfortable `claiming' the appreciation. This is because a contributor's \textit{sense of ownership} may not always correspond to objective measures of their impact in a collaborative effort. This is true of open source~\cite{young2021contributions, pinto2016more}, as well as Wikipedia~\cite{yim2024don}. We suggest that capturing appreciation in terms of abstractions of use can be a critical limitation in contexts where individual recognition is paramount.

\subsection{Encouraging appreciation in development practice}
Our approach departs from prior attempts to support appreciation in open source by connecting the appreciation system to the site where software is ultimately used---in the development environment. Deployment of our probe revealed many regularities in how appreciation was experienced and expressed by users. To the best of our knowledge, these findings are novel since prior studies tend to focus solely on the projects and contributors receiving appreciation~\cite{overney2020not, shimada2022github}. These findings also highlight new opportunities for encouraging appreciation. 

Our study revealed how participants felt and expressed appreciation in moments of transition, when they encountered new features, and in broader periods of reflection (\ref{moments}). Participants provided several ideas of how Hug Reports could explicitly account for these patterns, such as by scheduling interventions for when users are switching files or, on a broader time scale, when users are wrapping up a project (\ref{feedback}). Thus future designs can consider leveraging these moments to encourage reflection and expression of appreciation. 

Further, users proposed how different kinds of nudges could encourage them to express appreciation more often (\ref{feedback}). They suggested the system could indicate which packages they use most often, and could also indicate which packages have a smaller number of contributors. 

Even though contributors valued personal effort invested in the messages, many users struggled to find something specific to say in the moment (\ref{meanings}). Users and contributors both recognized the value of having added scaffolding to make personal effort more approachable (\ref{feedback}). The kinds of scaffolding suggested include predefined categories of \textit{thanks}, writing prompts, and even examples of thoughtful messages.

Across these proposals, however, it is important to note that even though reminders and writing support may encourage appreciation, like any design that lowers effort, they can also limit the meaningfulness of appreciation~\cite{wang2023reminders, liu2022will}. It is important to recognize what effort is interpreted as procedural, and what is interpreted as necessary for meaningful appreciation. Therefore, we suggest future work is necessary to evaluate the merit of these ideas.

\subsection{Limitations and future work}
\subsubsection{Limitations in the kinds of contributions considered}Our inquiry was limited to scenarios where open source software use occurred through packages. Future work can explore whether and how these approaches can be extended to other kinds of open source software such as end-user applications and cookie-cutter templates. Further, we chose to focus on Python and JavaScript packages because they are commonly used languages with large package ecosystems, and are also languages we are most familiar with. Future work can explore extending this approach to other programming languages/ecosystems. By relying on code activity traces, our work supported appreciation of contributors who made code contributions to projects. However, this approach overlooks many important non-code contributions such as documentation, governance, fund-raising, and community-building. Hence, there is an opportunity for future research to investigate how appreciation can be extended to non-code contributions. 

\subsubsection{Limitations in study methodology}
Our inquiry was heavily influenced by field deployment methods~\cite{siek2014field} and design methods~\cite{zimmerman2014research} intended to produce rich
qualitative accounts rather than statistically valid results. As with other field-based design research, our observations
are our own and other researchers working with the same problem framing may create other artifacts or pursue
different design activities, arriving at other, equally relevant conclusions~\cite{zimmerman2007research}. While our choice of design methods does not allow for statistical validity, we believe our work offers opportunities for what Zimmerman et al. describe as
“extensibility” [55]: that future attempts to develop technological support for appreciation, can build on our observations,
and our artifacts. Our research was also limited by the fact that users were encouraged to send \textit{thanks} at least two times every day they found themselves coding. This was to help ensure that each user would be able to try out the probe long enough to understand how its features influenced their interactions, and how it fit into their development practices, while also ensuring we would have a sufficient number of \textit{thanks} with which to study the experiences of contributors. This practice was consistent with other deployments of probes~\cite{leong2023social}, where similar to our study, research questions primarily concern participants' own descriptions and reflections rather than investigating voluntary adoption (\textit{``Will people use this tool?''}). This is also consistent with field studies that are concerned less with investigating voluntary adoption, which, as Siek et al. write~\cite{siek2014field}, feature``artificial inducements for adoption and use in order to focus on other factors such as the usefulness of specific system features, the appropriateness of the system in the given social context, the ability of the system to be appropriated for particular participant needs and practices, or the impacts of using the system on other factors, such as users’ behavior changes, work productivity, etc.'' Nevertheless, we recognize this decision limits the kinds of conclusions that can be drawn from our findings, and we follow the convention recommended by Siek et al.~\cite{siek2014field}, of reporting it in the study design, so that ``readers can carefully analyze the results in light of the compensation scheme''~\cite{siek2014field}.

\subsubsection{Studying the long-term impacts of appreciation} As a short-term technology probe deployment, our study could not investigate the long-term impacts of appreciation on motivation and participation in open source projects. Prior work identifies lack of recognition as one of the reasons due to which contributors disengage~\cite{guizani2021long, guizani2022attracting}. Hence, there is an opportunity for future work to investigate whether appreciation can increase retention. Further, prior work has also noted that exchanges that make an individual feel socially attached to a community can be effective at converting them into long-term contributors~\cite{kim2022receivers}. From this perspective, encouraging users to express appreciation has the potential to increase their social attachment to the project and eventually motivate them to contribute. Future work can explore whether and how exchanges of appreciation can create and engage a community around software artifacts. }

%% file: sections/conclusions.tex
\ed{Contributors of open source software packages rarely receive appreciation from users. In this paper, we observed how appreciation can be limited by the fact that \textit{where} users might feel appreciation (in their development environment) and \textit{what} they might feel appreciation towards (a package, its modules, or its functions) is detached from \textit{where} contribution activities occur (GitHub) and \textit{what} its units are (individual commits or pull requests). We described a field study of the Hug Reports \ed{technology probe} that provided users with a communication affordance within the code editor and allowed them to express appreciation in terms of the abstractions they are exposed to (package, modules). Our findings showed how Hug Reports encouraged appreciation in ways that were meaningful to users and contributors, how appreciation was interpreted both as a measure of utility and as an act of expressive communication, and that contributors' reactions to appreciation were influenced by how much they felt they had contributed to what was thanked. In addition to this, our study revealed patterns in when users expressed appreciation. We synthesized these findings into implications for developing appreciation systems in open source in particular, and peer production communities more generally.}

%% file: sections/appendix.tex
\section{Implementation Notes}
\label{implementationnotes}
\ed{
\subsection{Note on implementation details}To go from a concept proposal to a working technology probe, in addition to the key decisions of Hug Reports, we had to make choices for several low-level \textit{implementation details}. Examples of such choices include: (1) how often to notify contributors, and (2) whether the identities of senders or receivers should be revealed to each other. Many options were available for such choices and picking the \textit{best} option for each of these choices would warrant its own investigation. Determining the \textit{best} option for these choices was also orthogonal to answering our research questions, which evaluate the key decisions of Hug Reports and consider the cross-cutting requirements those key decisions reveal. In such situations, we often chose the option that was the simplest to implement within the constraints of the study. Our paper describes these choices for completeness and to contextualize our findings, while providing a reminder that other, potentially better, options are possible.
}

\subsection{Extension implementation}
The extension was written in TypeScript, using the VS Code API\footnote{https://code.visualstudio.com/api/references/vscode-api} and Contribution Points\footnote{https://code.visualstudio.com/api/references/contribution-points} for the core logic of the extension, and MongoDB Atlas\footnote{https://www.mongodb.com/atlas/database} to log interactions. First, all import statements in the code were detected using a regular expression to capture all possible forms in which a package, class, or function could be imported depending on the language (i.e. Python or JavaScript/Typescript).\\
(1) Regular expression to capture imports in Python: 
\begin{verbatim}
^(\s*(?:from\s+[\w\.]+)?\s*import\s+[\w\*\, ]+(?:\s+as\s+[\w]+)?)\b/gm
\end{verbatim}
(2) Regular expressions to capture imports in JavaScript/Typescript (either using the ``import'' or ``require'' keyword):
\begin{verbatim}
(a) /^import\s+.*\s+from\s+['"](.*)['"]/gm  
(b) /(const|let)\s+\{?\s*([\w,\s]+)\s*\}?\s*=\s*require\s*\(\s*['"]([^'"]+)['"]\s*\)[^;]*;/g
\end{verbatim}
From the lines that matched with these regular expressions, we extracted the names of the imported package, submodules, functions, and classes and stored these in an array called \inlinecode{python}{names}. We use \inlinecode{python}{name} to refer to each individual entry in the array. Beyond the import statements, to detect all lines of the file that interface with an external package, we had to identify lines that contained \inlinecode{python}{name.function()}, \inlinecode{python}{name()}, or \inlinecode{python}{name.attribute}. To do this, for each line in the file, we tested if the following patterns were present: \\
(1) To capture \inlinecode{python}{name.function()} and \inlinecode{python}{name()}: 
\begin{verbatim}
new RegExp(`\\b(?:${names.map(name => `(?:(?:${name})\\.\\w+|${name})`).join('|')})\\(`)
\end{verbatim}
(2) To capture \inlinecode{python}{name.attribute}:
\begin{verbatim}
new RegExp(`\\b(?:${names.map(name => `(?:(?:${name})\\.\\w+)`).join('|')})`)
\end{verbatim}
We stored a list of all line numbers at which a match was found and then, rendered the gutter icon using the \inlinecode{javascript}{setDecorations} function provided by the VS Code API on the set of lines where imports occurred and were used. For each rendered gutter icon, the ``Say Thanks'' option was configured by registering a menu contribution point as provided by the VS Code API. The additional modal pop-up to ``Say More'' was displayed using the \inlinecode{javascript}{showInformationMessage} function provided by the VS Code API.  

\section{Participant Table}
\label{bigger-participants}
\begin{table}[]
\caption{User demographics, their programming practices, and their current feelings and practices of expressing appreciation.}
\label{bigger-demo}
\centering
    \scriptsize
\begin{tabular}{lrlllll}
\hline
                     & \multicolumn{1}{l}{}                         &                 &                                                                                             &                                                                                                                                                       &                                                                                                                                                                                  &                                                                                                                                                                                                                                                                                                                                      \\
\textbf{Participant} & \multicolumn{1}{l}{\textbf{Age}}             & \textbf{Gender} & \textbf{\begin{tabular}[c]{@{}l@{}}Programming \\ proficiency\\ (self-report)\end{tabular}} & \textbf{\begin{tabular}[c]{@{}l@{}}Weekly hours spent,\\ on average, writing\\ Python, JavaScript,\\ or TypeScript code\\ (self-report)\end{tabular}} & \textbf{\begin{tabular}[c]{@{}l@{}}There are \\ many developers\\ whose work \\ I am grateful for \\ (seven point scale:\\ strongly disagree \\ to strongly agree)\end{tabular}} & \textbf{\begin{tabular}[c]{@{}l@{}}How often do you thank developers of open source projects \\ you use?\\ (open-ended question)\end{tabular}}                                                                                                                                                                                       \\
                     & \multicolumn{1}{l}{}                         &                 &                                                                                             &                                                                                                                                                       &                                                                                                                                                                                  &                                                                                                                                                                                                                                                                                                                                      \\ \hline
\rowcolor[HTML]{EFEFEF} 
                     & \multicolumn{1}{l}{\cellcolor[HTML]{EFEFEF}} &                 &                                                                                             &                                                                                                                                                       &                                                                                                                                                                                  &                                                                                                                                                                                                                                                                                                                                      \\
\rowcolor[HTML]{EFEFEF} 
U1                   & 23                                           & Man             & Advanced                                                                                    & 10                                                                                                                                                    & Agree                                                                                                                                                                            & Never                                                                                                                                                                                                                                                                                                                                \\
U2                   & 28                                           & Man             & Advanced                                                                                    & 10-15                                                                                                                                                 & Agree                                                                                                                                                                            & I hate to say this, but never :(                                                                                                                                                                                                                                                                                                     \\
\rowcolor[HTML]{EFEFEF} 
U3                   & 26                                           & Man             & Advanced                                                                                    & 5+                                                                                                                                                    & Agree                                                                                                                                                                            & \begin{tabular}[c]{@{}l@{}}Not often. If the projects are lesser-known, I would credit them in my\\ comments. However, if it's a widely used open-source project \\ (e.g., opencv, three.js, etc.), I don't tend to do it.\end{tabular}                                                                                              \\
U4                   & 25                                           & Woman           & Intermediate                                                                                & 15-20                                                                                                                                                 & Agree                                                                                                                                                                            & Almost never                                                                                                                                                                                                                                                                                                                         \\
\rowcolor[HTML]{EFEFEF} 
U5                   & 24                                           & Woman           & Advanced                                                                                    & 40                                                                                                                                                    & Agree                                                                                                                                                                            & Not enough :) We could do more in this domain.                                                                                                                                                                                                                                                                                       \\
U6                   & 28                                           & Woman           & Advanced                                                                                    & 20-30                                                                                                                                                 & Agree                                                                                                                                                                            & \begin{tabular}[c]{@{}l@{}}Not very often, unfortunately. I typically just download whatever NPM \\ package I need and feel more connected to the package and my feelings\\ towards the package (e.g., "Wow this documentation is good", "I don't \\ like how this API is designed", etc.) than towards the developers.\end{tabular} \\
\rowcolor[HTML]{EFEFEF} 
U7                   & 27                                           & Man             & Advanced                                                                                    & 8                                                                                                                                                     & Somewhat agree                                                                                                                                                                   & Almost never                                                                                                                                                                                                                                                                                                                         \\
U8                   & 25                                           & Man             & Intermediate                                                                                & 25                                                                                                                                                    & Agree                                                                                                                                                                            & \begin{tabular}[c]{@{}l@{}}There aren't many existing mechanisms to thank developers. Some \\ developers have a buy me a coffee link. For me, I sometimes \\ comment a thank you message but I don't specifically reach out to \\ the developers.\end{tabular}                                                                       \\
\rowcolor[HTML]{EFEFEF} 
U9                   & 26                                           & Woman           & Advanced                                                                                    & 5                                                                                                                                                     & Strongly agree                                                                                                                                                                   & 0 times ever                                                                                                                                                                                                                                                                                                                         \\
U10                  & 28                                           & Man             & Intermediate                                                                                & 20                                                                                                                                                    & Strongly agree                                                                                                                                                                   & \begin{tabular}[c]{@{}l@{}}Hardly ever.. I've only thanked people when I've met them in person \\ at conferences and such..\end{tabular}                                                                                                                                                                                             \\
\rowcolor[HTML]{EFEFEF} 
U11                  & 29                                           & Man             & Advanced                                                                                    & 7                                                                                                                                                     & Agree                                                                                                                                                                            & \begin{tabular}[c]{@{}l@{}}I've chatted with other developers in Discord spaces and through \\ GitHub pull requests. My best answer would be sporadically - \\ for anything during the interaction. Rarely for an "overall" thanks\\ for the body of work, or for the overall project.\end{tabular}                                  \\
U12                  & 25                                           & Man             & Expert                                                                                      & 40                                                                                                                                                    & Strongly agree                                                                                                                                                                   & Never                                                                                                                                                                                                                                                                                                                                \\
\rowcolor[HTML]{EFEFEF} 
U13                  & 24                                           & Woman           & Advanced                                                                                    & 10                                                                                                                                                    & Agree                                                                                                                                                                            & Very rarely, only if I know them personally                                                                                                                                                                                                                                                                                          \\
U14                  & 24                                           & Man             & Advanced                                                                                    & 28                                                                                                                                                    & Agree                                                                                                                                                                            & Not often                                                                                                                                                                                                                                                                                                                            \\
\rowcolor[HTML]{EFEFEF} 
U15                  & 23                                           & Man             & Advanced                                                                                    & 10                                                                                                                                                    & Strongly agree                                                                                                                                                                   & \begin{tabular}[c]{@{}l@{}}I've never thanked someone explicitly, unless I happen to meet them \\ in person, but I do often star useful or interesting projects on GitHub, \\ and will suggest relevant projects to my peers.\end{tabular}                                                                                           \\
U16                  & 30                                           & Man             & Intermediate                                                                                & 45                                                                                                                                                    & Strongly agree                                                                                                                                                                   & Rarely                                                                                                                                                                                                                                                                                                                               \\
\rowcolor[HTML]{EFEFEF} 
U17                  & 27                                           & Woman           & Advanced                                                                                    & 20                                                                                                                                                    & Strongly agree                                                                                                                                                                   & \begin{tabular}[c]{@{}l@{}}I have never explicitly expressed my gratitude to the developers of open \\ source projects unless they're friends of mine.\end{tabular}                                                                                                                                                                  \\
U18                  & 26                                           & Man             & Advanced                                                                                    & 20                                                                                                                                                    & Strongly agree                                                                                                                                                                   & \begin{tabular}[c]{@{}l@{}}I think I used to thank developers a lot more when I was still active \\ in the open source space with {[}project{]}. But haven't done so any time \\ recently.\end{tabular}                                                                                                                              \\
                     & \multicolumn{1}{l}{}                         &                 &                                                                                             &                                                                                                                                                       &                                                                                                                                                                                  &                                                                                                                                                                                                                                                                                                                                      \\ \hline
\end{tabular}
\end{table}